\documentclass{article}
\usepackage{graphicx}  
\usepackage{amsmath}   
\usepackage[compress]{cite}
\usepackage{amssymb}   
\usepackage{bm} 
\usepackage{dcolumn}
\usepackage{color}
\usepackage{xcolor}
\usepackage{mathrsfs}
\usepackage{amsfonts}
\usepackage{varioref}
\usepackage{textcomp}
\usepackage[normalem]{ulem}

\usepackage{multirow}
\usepackage{caption}
\usepackage{subcaption}
\RequirePackage[colorlinks,citecolor=blue,urlcolor=magenta,linkcolor=blue]{hyperref}
\allowdisplaybreaks

\labelformat{section}{Section #1} 
\labelformat{subsection}{Section #1} 
\labelformat{subsubsection}{Section #1}
\labelformat{subsubsubsection}{Section #1}
\labelformat{equation}{Eq.~(#1)} 
\labelformat{figure}{Fig.~#1} 
\labelformat{subfigure}{Fig.~\thefigure#1} 
\labelformat{table}{Table~#1} 
\labelformat{appendix}{Appendix #1}

\title{\bf Tidal deformation of black holes in Lovelock gravity}

\author{Chiranjeeb Singha\footnote{chiranjeeb.singha@iucaa.in}$~^{1}$ and Sumanta Chakraborty\footnote{tpsc@iacs.res.in}$~^{2}$\\
$^{1}$\small{Inter-University Centre for Astronomy \& Astrophysics, Post Bag 4, Pune 411 007, India}\\
$^{2}$\small{ School of Physical Sciences, Indian Association for the Cultivation of Science, Kolkata 70032, India}}

\begin{document}
  
\maketitle
\begin{abstract} 
It is well established that black holes in four-dimensional, vacuum, general relativity exhibit vanishing static tidal Love numbers, indicating no multipolar response to the external tidal fields in the static limit. This intriguing feature does not extend to higher-dimensional spacetimes within general relativity, where static black holes can possess non-zero static tidal Love numbers (TLNs). In this work, we have examined the tidal deformation of black holes in Lovelock gravity. We find that, in certain cases within pure Lovelock gravity, the static TLN vanishes, extending the four-dimensional result to specific higher-dimensional settings. On the other hand, black holes in Einstein–Gauss–Bonnet gravity consistently exhibit non-zero static TLNs, with their magnitude depending on the Gauss–Bonnet coupling constant. Exceptions occur only in certain special cases associated with axial perturbations. These results highlight the sensitivity of the multipolar response of a black hole under tidal field, to the underlying theory of gravity and the spacetime dimensions.
\end{abstract}
\section{Introduction}

Despite their deeply mysterious nature, owing to their behaviour as a one-way membrane, black holes (BHs) appear remarkably simple to an external observer. The uniqueness of the Kerr-Newman solution in general relativity (GR) \cite{Kerr:1963ud, Newman:1965my}, encapsulated in the celebrated no-hair theorems \cite{Israel:1967wq, Carter:1971zc, Bekenstein:1971hc}, implies that BHs are among the simplest objects in the universe, characterized entirely by their mass, charge, and spin\footnote{The classic no-hair theorems can be extended to include more exotic forms of matter \cite{Bekenstein:1995un, Hui:2012qt} under static boundary conditions. Interestingly, black holes can support time-dependent scalar profiles with non-trivial boundary conditions \cite{Jacobson:1999vr, Horbatsch:2011ye, Hui:2019aqm, Clough:2019jpm}, as well as superradiant clouds \cite{Penrose:1969pc, 1971JETPL..14..180Z, 1972JETP...35.1085Z, 1972BAPS...17..472M, Bardeen:1972fi, Press:1972zz, Starobinsky:1973aij, Teukolsky:1974yv}; see \cite{Arvanitaki:2010sy, Endlich:2016jgc, Baumann:2019eav} for recent discussions. However, we will concentrate on the static case, and hence no-hair theorems will hold good here.}. Although a small set of parameters describes the BH solutions, it is natural to ask how they respond to external perturbations, such as an external tidal perturbation. These responses are intrinsic quantities that further characterize a BH and are worth understanding. The study of tidal responses of various compact objects is not merely academic; these effects contribute to gravitational wave (GW) emission and are in principle measurable \cite{Cardoso:2017cfl, Chakraborty:2025wvs}\footnote{For neutron stars, the tidal deformations serve as key observables, as their values can be quite large $\sim \mathcal{O}(10^{2})$\cite{Hinderer:2007mb,  hinderer2009tidal-327} and hence detectable by the present generation of GW detectors. These deformations play a crucial role in constraining the equation of states of neutron stars~\cite{yagi2013multipole-5d7, Yagi:2016bkt, LIGOScientific:2017vwq}.}.

The linear tidal response of a compact object due to an external tidal perturbation can be obtained through linear perturbation theory. As an example, consider the quadrupolar, static tidal perturbations $\mathcal{E}_{ij}$ to static and spherically symmetric spacetimes, under which, the compact objects will generate a tidal moment $Q_{ij}$, related to the tidal field in the linear perturbation theory as, $Q_{ij}=-\lambda_{2} \mathcal{E}_{ij}$. Here, $\lambda_{2}$ measures the tidal deformability associated with the quadrupolar gravitational perturbations and is a dimensionful quantity. It is possible to construct a dimensionless quantity $k_{2}$, given by $\lambda_{2}=(2/3)k_{2} R^{5}$, which is referred to as the quadrupolar tidal Love number (TLN) \cite{Hinderer:2007mb, Binnington:2009bb, Damour:2009vw}, where $R$ is the radius of the compact object. For generic multipoles in a spherically symmetric background, the TLNs are denoted by $k_{\ell}$, with $\ell=2$ being a quadrupolar perturbation. The TLNs, so defined, are for static configurations and hence are referred to as static TLNs, appearing in the GW signal at a relative fifth post-Newtonian order \cite{hinderer2009tidal-327}. The determination of the TLNs in the Newtonian limit follows from the asymptotic expansion of the Newtonian potential in the presence of a tidal field, $U=\alpha r^{2}+\cdots+(\beta/r^{3})+\cdots$, and hence the quadrupolar TLN is given by $k_{2}\propto (\beta/\alpha)$. In the relativistic context, on the other hand, one uses linear perturbation of gravitational field equations to compute the TLN of compact objects, where one uses appropriate master functions constructed out of metric perturbations and studies their asymptotic behaviour. Just as the metric perturbations can be divided into axial (transforming as $(-1)^{\ell+1}$ under parity) and polar (transforming as $(-1)^{\ell}$ under parity) sectors, there are also axial as well as polar TLNs. The polar TLNs, which have a well-defined Newtonian limit, are obtained by taking the ratio of the coefficient of the $r^{-\ell-1}$ term with the coefficient of the $r^{\ell}$ term. Similarly, the axial TLNs are obtained from the ratio of the coefficient of $r^{-\ell}$ with the coefficient of $r^{\ell+1}$. The analysis presented above is for four-dimensional spacetime, which can be generalized to higher-dimensional static and spherically spacetimes in a straightforward manner, in which, in addition to polar and axial sectors, there is a tensor sector in the metric perturbations \cite{Takahashi:2010ye, Ishibashi:2003ap, Hui:2020xxx}. The TLNs of a higher dimensional Schwarzschild BH under all possible perturbations have already been calculated in \cite{Hui:2020xxx}, while the TLNs of a higher dimensional Kerr BH (Myers-Perry solution) have been calculated only for scalar perturbations in \cite{charalambous2023scalar-ba1}.

The TLNs, as defined above, were first studied for spherically symmetric objects in \cite{Hinderer:2007mb, Binnington:2009bb, Damour:2009vw}, revealing intriguing properties. Notably, the axial and polar TLNs of a four-dimensional Schwarzschild BH vanish identically \cite{Damour:2009vw, Binnington:2009bb, Fang:2005qq, Kol:2011vg, Chakrabarti:2013lua, Gurlebeck:2015xpa}, suggesting that the gravitational potential of static and spherically symmetric vacuum configurations in GR cannot be deformed\footnote{The above result can be arrived at from various possible directions, e.g., existence of a ladder-like symmetry can be argued for the existence of zero TLNs \cite{hui2022ladder-678, achour2022hidden-8c7, charalambous2021hidden-5e0}. Similarly, from the matching of BH perturbation theory with the effective field theory computation or the relevant scattering amplitude, one again arrives at the fact that BHs in four-dimensional GR have vanishing TLNs \cite{ivanov2023vanishing-9aa, creci2021tidal-42e, Charalambous:2021mea}.}. However, this vanishing of the TLNs is not generic to all BHs, e.g., the TLNs of Schwarzschild BH do not vanish in higher dimensions \cite{Kol:2011vg, Cardoso:2019vof, Hui:2020xxx}, with anti-de Sitter asymptotics \cite{Emparan:2017qxd}, in the presence of higher-curvature terms \cite{Cardoso:2018ptl}, in many alternative theories of gravity \cite{Cardoso:2017cfl, DeLuca:2022tkm}, for Schwarzschild-de Sitter BHs \cite{Nair:2024mya}, for area-quantized BHs \cite{Nair:2022xfm}, for the BHs in $(2+1)$ dimensions \cite{DeLuca:2024ufn, Bhatt:2024mvr} and in dynamical scenario \cite{Chakraborty:2025wvs, Bhatt:2023zsy, Bhatt:2024yyz, katagiri2024relativistic-5c0, creci2021tidal-42e, saketh2023dynamical-b30, Chakraborty:2023zed, Kehagias:2024rtz}. This indicates that their vanishing is a special feature of static perturbations of general relativity in four dimensions\footnote{For Kerr BHs, as well the static TLNs vanish in four dimensions \cite{LeTiec:2020bos, LeTiec:2020spy, Bhatt:2023zsy, Gounis:2024hcm}. This was first shown for slowly spinning BHs under axisymmetric perturbations in \cite{Pani:2015hfa, Pani:2015nua, Landry:2015zfa, Landry:2015cva, Landry:2017piv, Poisson:2020mdi} and later extended to general spins \cite{LeTiec:2020spy, LeTiec:2020bos, Chia:2020yla, Bhatt:2024rpx, Bhatt:2024yyz}. However, Kerr BHs exhibit dissipative tidal heating due to their rotation \cite{Goldberger:2020fot, Chia:2020yla, Charalambous:2021mea, Bhatt:2023zsy}, even in the static limit.}.

In this paper we ask the question, what are the alternative theories of gravity with second-order field equations, thereby evading ghosts? The answer to this question is well known and holds for gravity theories broadly divided into two classes---(a) the scalar tensor theories, e.g., Brans-Dicke and Horndeski \cite{Kobayashi:2019hrl} and (b) higher curvature theories, e.g., Lovelock theories of gravity \cite{Padmanabhan:2013xyr}. While TLNs in Horndeski theories of gravity have received recent attention \cite{Diedrichs:2025vhv, Creci:2023cfx, Creci:2024wfu}, the TLNs of the Lovelock class of theories have remained unattended. To fill this gap we will be studying the TLNs of BHs in the pure Lovelock theories of gravity and also in Einstein-Gauss-Bonnet gravity in this paper. Pure Lovelock theories of gravity exist in dimensions greater than four and with a Lagrangian consisting of a polynomial of order $N$ in the Riemann curvature tensor. When $N=1$, the theory reduces to a linear term corresponding to the scalar curvature, recovering Einstein's theory. Generally, the spacetime dimension $d$ and the Lovelock order $N$ are independent, with the constraint that $N>1$ is allowed only for $d>4$. Additionally, for the existence of non-trivial solutions to the gravitational field equations, we must have $d\geq 2N+2$ \cite{Gannouji:2019gnb, Dadhich:2015nua, Dadhich:2012cv, Gannouji:2013eka, Dadhich:2015ivt, Padmanabhan:2013xyr, Dadhich:2015lra}. Static and spherically symmetric BHs in pure Lovelock theories are well known \cite{Dadhich:2015nua, Dadhich:2012cv, Gannouji:2013eka, Dadhich:2015ivt} and hence the study of their perturbations leading to TLNs will be the primary focus of this work. The Einstein–Gauss–Bonnet (EGB) gravity, on the other hand, represents the simplest nontrivial extension of general relativity within Lovelock’s generalization involving the two terms of the Lovelock polynomial, $N=1$ (the Einstein term) and $N=2$ (the Gauss-Bonnet term) \cite{Lovelock:1971yv}. Static and spherically symmetric BH in EGB gravity is also well known and is given by the Boulware-Deser solution \cite{boulware1985string-generated-4e7, garraffo2008lovelock-70a}. We will also study its perturbation and hence determine the associated static TLNs.

This paper is arranged as follows: In \ref{lovelock}, we will consider the scalar, tensor, axial, and polar perturbations of a static and spherically symmetric BH in the pure Lovelock theories of gravity. Using these perturbations, we will calculate the static TLNs of these BHs in pure Lovelock gravity. Subsequently, in \ref{Gauss}, we will consider the scalar, tensor, axial, and polar perturbations for the Boulware-Deser BH in the EGB gravity, and will calculate the static TLNs of the Boulware-Deser BH. Finally, we will conclude in \ref{conclusion}, highlighting our results as well as presenting potential avenues for future research. \ref{abcd} presents the solutions to the hypergeometric equation relevant for each of the perturbations. In \ref{App:gravlove}, we derive the algebraic equation from which the solutions to the gravitational field equations in Lovelock gravity can be obtained.

{\em Notations and conventions}: In this paper, we adopt the positive signature convention, where the Minkowski metric in $(3+1)$ dimensions expressed in Cartesian coordinates is given by: $\text{diag.}(-1, +1, +1, +1)$. Additionally, we set $G = c = 1$ throughout the paper. The Greek indices $\mu,\nu,\rho,\cdots$, denote spacetime coordinates, while Roman indices $i,j,a,\cdots$, denote spatial coordinates. 
\section{Static tidal Love numbers for black holes in pure Lovelock theories of gravity}\label{lovelock}

In this section, we will analyze the scalar and gravitational perturbations of static and spherically symmetric BHs in pure Lovelock theories of gravity. The EM perturbation and hence the EM TLNs can be obtained in a similar manner, which we will not present here. In the static situation, the corresponding perturbation equations can be solved analytically, and from the asymptotic behaviour of these perturbations, we will determine the associated static TLNs. Note that we divide the gravitational perturbations into three classes --- (a) axial, (b) polar and (c) tensor, following \cite{Takahashi:2010ye}. Thus we have consciously divided the perturbations into four groups: (a) scalar, (b) tensor, (c) axial and (d) polar perturbations. In what follows we discuss each of them separately.

To start with, we present below the static and spherically symmetric BH solution in pure Lovelock theories. This is obtained by solving the vacuum gravitational field equations of pure Lovelock theories of order $N$ in $d$ spacetime dimensions ($d\geq2N+2$), arising from the following Lagrangian
\begin{equation}\label{Lovelock_Lagrangian}
L_{\rm N}=\frac{1}{2^{N}}\delta^{\mu_{1}\mu_{2}\cdots \mu_{2N-1}\mu_{2N}}_{\nu_{1}\nu_{2}\cdots \nu_{2N-1}\nu_{2N}}R^{\nu_{1}\nu_{2}}_{\mu_{1}\mu_{2}}\cdots R^{\mu_{2N-1}\mu_{2N}}_{\nu_{2N-1}\nu_{2N}}~,
\end{equation}
where, $R^{\mu\nu}_{\alpha\beta}$ is the Riemann curvature tensor, $\delta^{\mu\nu\cdots}_{\alpha\beta\cdots}$ is the completely antisymmetric determinant tensor, such that $\delta^{\mu \nu}_{\alpha \beta}=(\delta^{\mu}_{\alpha}\delta^{\nu}_{\beta}-\delta^{\mu}_{\beta}\delta^{\nu}_{\alpha})$. The metric for the static and spherically symmetric BH arising from the above Lagrangian reads \cite{Gannouji:2019gnb, Dadhich:2015nua, Dadhich:2012cv, Gannouji:2013eka, Dadhich:2015ivt, Padmanabhan:2013xyr, Dadhich:2015lra},
\begin{equation}\label{pure_love_BH}
ds^{2}=-f(r)dt^{2}+f(r)^{-1} dr^{2}+r^{2}d\Omega_{d-2}^{2}~;
\qquad 
f(r)=1-\left(\frac{2^{N}M}{r^{d-2N-1}}\right)^{\frac{1}{N}}~,
\end{equation}
where $M$ is the ADM mass of the BH and $d\Omega_{d-2}^{2}$ is the line element on the $(d-2)$-dimensional sphere. Note that, the horizon of the BH is located at $r_{\rm H}$, which is the largest root of the equation: $r_{\rm H}^{d-2N-1}=2^{N}M$. Using this, one can also express the metric function $f(r)$ as, $f(r)= 1-\left(r_{\rm H}/r\right)^{\alpha}$, with $\alpha\equiv\{(d-2N-1)/N\}$. We will use this particular parametrization of the metric components of the pure Lovelock BH in the subsequent discussion.
\subsection{Scalar Love numbers for black holes in pure Lovelock gravity}

Having described the unperturbed background, we delve into the discussion of TLNs. As a warm up, we start with the scalar perturbation and the determination of scalar TLNs for pure Lovelock BH, described by \ref{pure_love_BH}. For this purpose, we consider a massless scalar field, living on this background, satisfying the Klein-Gordon equation $ g^{\mu\nu} \nabla_{\mu}\nabla_{\nu} \Phi=0$. Given that the background metric is static and spherically symmetric, it follows that, the scalar field can be decomposed as,
\begin{equation}\label{}
\Phi(t,r,\Omega) = \sum_{\ell,m}e^{-i\omega t}\frac{\phi_{\ell m}(r)}{r^{(d-2)/2}}\mathcal{Y}_{\ell \mathbf{m}}(\Omega)~,
\end{equation}
where, $\mathcal{Y}_{\ell \mathbf{m}}(\Omega)$ are spherical harmonics on a $(d-2)$-dimensional sphere, used to describe the angular dependence of the solutions to the field equations. Here, $\mathbf{m}=(m_{1},m_{2},\cdots,m_{d-3})$ are the quantum numbers associated with all the $(d-3)$ angles $(\theta_{1},\theta_{2},\cdots,\theta_{d-3})$, and $\ell$ is the quantum number associated with $\theta_{d-2}$, such that $\Omega=(\theta_{1},\theta_{2},\cdots,\theta_{d-3},\theta_{d-2})$. The spherical harmonic function $\mathcal{Y}_{\ell \mathbf{m}}(\Omega)$ satisfies the following differential equation: $D_{A}D^{A}\mathcal{Y}_{\ell \mathbf{m}}(\Omega)=-\ell(\ell+d-3)\mathcal{Y}_{\ell \mathbf{m}}(\Omega)$, where $D_{A}$ is the covariant derivative on the $(d-2)$-sphere. The radial function $\phi_{\ell m}(r)$, on the other hand, satisfies the following differential equation,
\begin{eqnarray}
&&\left(\frac{\partial^2}{\partial r_{*}^2}+\omega^2-V^{(0)}_{\rm eff}(r)\right)\phi_{\ell m}(r)=0
\nonumber
\\
&&V_{\rm eff}^{(0)}(r)= f(r)\left\{\frac{\ell(\ell+d-3)}{r^2}+\frac{(d-2)(d-4)}{4r^2}f(r)+\frac{(d-2)}{2r}f'(r)\right\}~,
\end{eqnarray}
where $dr_{*}=\{dr/f(r)\}$ is the tortoise coordinate and $V_{\rm eff}^{(0)}(r)$ is the effective potential experienced by the scalar field. For the static case ($\omega=0$), we arrive at,
\begin{equation}\label{eqn4}
f\frac{\partial^{2}\phi_{\ell m}}{\partial r^{2}}+f'\frac{\partial\phi_{\ell m}}{\partial r}
-\left\{\frac{\ell(\ell+d-3)}{r^2}+\frac{(d-2)(d-4)}{4r^2}f(r)+\frac{(d-2)}{2r}f'(r)\right\}\phi_{\ell m}=0~.
\end{equation}
The above equation can be casted in a familiar form, by introducing a new radial coordinate, $x\equiv\left(r_{\rm H}/r\right)^{\alpha}$, with $\alpha\equiv\{(d-2N-1)/N\}$ and a rescaled radial part of the scalar field: $u(x)\equiv x^{-\beta}\phi_{\ell m}(x)$. The new radial coordinate is defined such that the horizon is characterized by, $x=1$, while $x=0$ corresponds to $r\to \infty$. In terms of these redefined coordinates and rescaled scalar field, \ref{eqn4} becomes,
\begin{align}\label{eqn5}
x(1-x)u''(x)&+\bigg\{c-(a+b+1)x\bigg\}u'(x)-a b u(x)=0~,
\\
c=\frac{1}{\alpha}+1+2\beta&=a+b\,;
\qquad 
ab=\frac{\beta}{\alpha}(1+\alpha \beta)+\beta+\frac{(d-2)}{2 \alpha}-\frac{(d-4)(d-2)}{4 \alpha^{2}}
\nonumber
\end{align}
provided, the quantity $\beta$, present in the rescaled radial perturbation of the scalar field, is obtained from the following algebraic equation,
\begin{equation}
\alpha\beta\left(1+\alpha\beta\right)-\ell(\ell+d-3)-\frac{(d-4)(d-2)}{4}=0\,,
\end{equation}
which has the following solution: $\alpha\beta=\{(d+2\ell-4)/2\}$\footnote{As the above equation is quadratic, there will be two solutions, one of which we have presented here. The other solution has been discarded, since it does not have a well-behaved general relativistic limit \cite{Hui:2020xxx}.}. Given $\alpha=\{(d-2N-1)/N\}$, it follows that, $\beta=(N/2)(d+2\ell-4)(d-2N-1)$. These choices for $\alpha$ and $\beta$ uniquely fix the constants $a$ and $b$, appearing in the hypergeometric differential equation. Therefore, we obtain the following solutions for the quantities $a$, $b$ and $c$ as,
\begin{equation}\label{abc}
\left.
\begin{array}{c}
a=\hat{\ell}+\hat{d}
\\
b=1+\hat{\ell}
\\
c=1+2\hat{\ell}+\hat{d}
\end{array}
\right\}
\qquad 
\hat{\ell}\equiv\frac{\ell}{\alpha}~;
\quad
\hat{d}\equiv\frac{(d-3)}{\alpha}~,
\end{equation}
where, we have introduced the quantities $\hat{\ell}$ and $\hat{d}$ for convenience. In this case, $c=a+b$, and hence all the results of \ref{abcd} become applicable. For $N=1$, i.e., GR, it follows that $\hat{d}=1$ and $\hat{\ell}=\ell/(d-3)$, recovering the results of \cite{Hui:2020xxx} for $d$ dimensional Schwarzschild BH. 

The pure Lovelock theory of order $N$ is non-dynamical in $d=2N+1$, and hence for non-trivial dynamics, one must consider $d=2N+1+n$, where $n\in \mathbb{Z}^{+}$. Then, the quantities $(a,b,c)$, appearing in the hypergeometric equation becomes, 
\begin{equation} \label{ceqn}
\left.
\begin{array}{c}
a=\frac{N}{n}(\ell+d-3)=N+\frac{N}{n}(\ell+2N-2)
\\
b=1+\frac{N}{n}\ell\,;
\\
c=\frac{2N (\ell+N-1)}{n}+(N+1)
\end{array}
\right\}\,.
\end{equation}
As evident, for generic choices of the spacetime dimension $d$, Lovelock order $N$ and angular harmonic $\ell$, the entries of the hypergeometric function $(a,b,c)$ take generic values, and are not integers. Of course, there can be specific cases in which $(a,b)$ are half-integers with $c$ being an integer, or $(a,b,c)$ are all integers. Given this setting and the above possible choices of $(a,b,c)$, presented in \ref{abc}, the static TLNs of pure Lovelock BHs under scalar perturbation can be determined as:

\begin{itemize}

\item \textbf{Case 1:}  For generic values of $a$, $b$, and $c$, the radial part of the scalar perturbation $\Phi$ can be written as, $\Phi_{\ell \mathbf{m}}(r)\equiv r^{-(d-2)/2}\phi_{\ell \mathbf{m}}(r)=r^{-(d-2)/2}(r_{\rm H}/r)^{\alpha\beta}u(r)$, which has the following asymptotic behaviour (see \ref{response_gen} of \ref{abcd}),
\begin{equation}
\Phi^{\rm (gen)}_{\ell \mathbf{m}}(r)=\mathcal{A}(-2\hat{\ell}-\hat{d})\frac{\Gamma(2\hat{\ell}+\hat{d})}{\Gamma(\hat{\ell}+\hat{d})\Gamma(\hat{\ell}+1)}r^{-\frac{(d-2)}{2}}\left(\frac{r_{\rm H}}{r}\right)^{-2\ell-(d-3)+\alpha \beta}\left[1+\frac{\mathcal{F}^{\rm (I)}r_{\rm H}^{2\ell+d-3}}{r^{2\ell+d-3}}\right]\,.
\end{equation}
Therefore, the response function is given by $\mathcal{F}^{\rm (I)}$, and has the expression given in \ref{response_I} of \ref{abcd}. Therefore, the TLNs become,
\begin{equation}
k^{\rm scalar\,(gen)}_{\ell}=\frac{1}{2}\textrm{Re}\left(\mathcal{F}^{\rm (I)}\right)
=\frac{(2\hat{\ell}+\hat{d})}{2}
\frac{\Gamma(\hat{\ell}+\hat{d})^2 \Gamma(\hat{\ell}+1)^2}{\Gamma(1+2\hat{\ell}+\hat{d})^2}
\frac{\sin[\pi (\hat{\ell}+\hat{d})]\sin[\pi (\hat{\ell}+1)]}{\pi \sin[\pi(2\hat{\ell}+\hat{d}+1)]}\,.
\label{generallove32}
\end{equation}
Note that the TLNs are dimensionless and are related to the dimensionful deformation parameters $\lambda_{\ell}$ as, $\lambda_{\ell}=k_{\ell}r_{\rm H}^{-2\ell-(d-3)}$. There are instances, where the normalization factor is $M^{-2\ell-(d-3)}$, and will differ from the above TLNs by an overall factor of $(M/r_{\rm H})^{2\ell+d-3}$.

\item \textbf{Case 2:} For $a$ and $b$ being half integers and $c$ is an integer, the radial part of the scalar perturbation $\Phi$ can be expressed as in the case of arbitrary $(a,b,c)$, however has a different asymptotic behaviour (see \ref{response_half} of \ref{abcd}),
\begin{align}
\Phi^{\rm (mix)}_{\ell \mathbf{m}}(r)&=B(-1)^{2 \hat{\ell}+\hat{d}+1}(2 \hat{\ell}+\hat{d}) \frac{\Gamma(2 \hat{\ell}+\hat{d})^2}{\Gamma(\hat{\ell}+\hat{d})^2\Gamma(\hat{\ell}+1)^2}\frac{\pi^{2}}{\sin[\pi (\hat{\ell}+\hat{d})]\sin[\pi (\hat{\ell}+1)]}r^{-\frac{(d-2)}{2}}
\nonumber
\\ 
&\times\left(\frac{r_{\rm H}}{r}\right)^{-2\ell-(d-3)+\alpha \beta}\left[1+\frac{\mathcal{F}^{\rm (II)}r_{\rm H}^{2\ell+d-3}}{r^{2\ell+d-3}}\right]\,.
\end{align}
Therefore, the response function and hence the TLNs become,
\begin{align}\label{scalarmixed}
k^{\rm scalar\,(mixed)}_{\ell}&=\frac{1}{2}\textrm{Re}\left(\mathcal{F}^{\rm (II)}\right)
\nonumber
\\
&=(-1)^{2\hat{\ell}+\hat{d}+1}\bigg(\frac{\alpha(2\hat{\ell}+\hat{d})}{2}\bigg)\frac{\Gamma(\hat{\ell}+\hat{d})^2 \Gamma(\hat{\ell}+1)^2}{\Gamma(2\hat{\ell}+\hat{d})^2}
\frac{\sin[\pi (\hat{\ell}+\hat{d})]\sin[\pi (\hat{\ell}+1)]}{\pi^{2}}\ln\bigg(\frac{r_{\rm H}}{r}\bigg)~.
\end{align}
We have used the same normalization for the TLNs here as well, i.e., the TLNs are normalized by the horizon radius, rather than the mass of the pure Lovelock BH. 

\textit{Example:} Consider the following relation between spacetime dimension and Lovelock order: $d=2N+3$, which implies $n=2$. In this case, if we further assume both $N$ and $\ell$ to be odd, then $c$ is an integer, while $a$ and $b$ are half-integers.

\item \textbf{Case 3:} For integer values of $a$, $b$, and $c$, the radial part of the scalar perturbation $\Phi$ has the following asymptotic behaviour (see \ref{response_Int} of \ref{abcd}),
\begin{equation}
\Phi^{\rm (int)}_{\ell \mathbf{m}}(r)=-A\frac{\Gamma(1+2\hat{\ell}+\hat{d})}{\Gamma(\hat{\ell}+\hat{d})\Gamma(\hat{\ell}+1)}r^{-\frac{(d-2)}{2}}\left(\frac{r_{\rm H}}{r}\right)^{-2\ell-(d-3)+\alpha \beta}\,;\qquad 
\mathcal{F}^{\rm (III)}=0\,.
\end{equation}
As evident, in this case the scalar field has no decaying mode, and hence it follows that the TLNs become,
\begin{equation}
k^{\rm scalar\,(int)}_{\ell}=\frac{1}{2}\textrm{Re}\left(\mathcal{F}^{\rm (III)}\right)=0~.
\end{equation}
A corollary of the above result being, for $N=1$, we have general relativity, for which the TLNs of vacuum BH in $d=4$ vanish. Another interesting fact arising out of the above expression is the result that vacuum BHs in spacetime dimension $d=3N+1$, with arbitrary Lovelock order $N$, have vanishing scalar TLNs. 

\textit{Example:} (a) Consider the case in which the quantity $(N/n)$ is an integer, implying $N=m\times n$, with $(m,n)\in \mathbb{Z}^{+}$. For this choice, it follows from \ref{ceqn} that $a$, $b$, and $c$ are all integers. The choices $d=3N+1$ as well as $d=2N+2$ are special cases of this scenario, for which $(a,b,c)$ are all integers. (b) Consider $n=2$, which implies $d=2N+3$. In this case for (i) even $N$ or (ii) odd $N$ and even $\ell$, $a$, $b$, and $c$ are integers.
 
\end{itemize}

The above summarizes the TLNs of vacuum BHs in spacetime dimension $d$, arising from pure Lovelock theories of order $N$, under scalar perturbation. In the subsequent sections we will discuss the TLNs associated with different irreducible parts of the gravitational perturbation. 

\subsection{Static tidal Love numbers of black holes in pure Lovelock gravity under tensor perturbation}

Gravitational perturbation of higher dimensional static and spherically symmetric BH spacetimes can be decomposed into three irreducible parts --- (a) scalar perturbation, which transforms as $(-1)^{\ell}$ under parity, and is referred to as polar perturbations in four dimensions; (b) vector perturbation, which is parity odd, and is the axial perturbation in four dimensions; and finally (c) tensor perturbation, which has no analog in four spacetime dimensions. Due to the simplicity, we will first discuss the tensor perturbation, before moving to discuss the axial and then finally the polar perturbation. 

The study of gravitational perturbation in Lovelock theories of gravity crucially hinges on the symmetries of the background spacetime, as well as properties of the alternating tensor, presented in \ref{Lovelock_Lagrangian}. For the purpose of the reader, we have summarized the basic techniques used to determine the background geometry as well as the perturbations in \ref{App:gravlove}. Following which, it follows that, in pure Lovelock gravity, the following quantities are important in determining the perturbations:
\begin{align}
T(r)&=Nr^{(d-2N-1)/N}=Nr^{\alpha}\,;
\qquad 
\frac{T'}{T}=\frac{d-2N-1}{Nr}=\frac{\alpha}{r}\,,
\label{T1}
\\
\frac{T''}{T'}&=\frac{d-3 N -1}{N r}=\frac{\alpha-1}{r}\,;
\qquad 
\frac{T'''}{T'}=\frac{(\alpha-1)(\alpha-2)}{r^{2}}\,,
\label{T2}
\end{align}
as they enter the master equation satisfied by the tensor perturbations through the effective potential, which reads \cite{Takahashi:2010ye}:
\begin{align}
\bigg\{\frac{\partial^2}{\partial r_{*}^2}&+\omega^2-V_{\rm T}(r)\bigg\}\Psi_{\rm T}(r)=0\,,
\\
V_{\rm T}(r)&=\frac{\ell(\ell+d-3)f(r)}{(d-4)r}\left(\frac{d \ln{T^{'}}}{dr}\right)+\frac{1}{r\sqrt{T^{'}}}f\frac{d}{dr}\left(f\frac{d}{dr}r\sqrt{T^{'}}\right)\,,
\label{tensor}
\end{align}
where the radial function $f(r)$ is given by \ref{pure_love_BH} and $\Psi_{\rm T}(r)$ is the tensor master function. Here, $\alpha=\{(d-2N-1)/N\}$, which has been defined earlier. Using the expressions for $T(r)$ and its derivatives from \ref{T1} and \ref{T2}, we get the radial master equation for static case ($\omega=0$) in the following form,
\begin{equation}\label{tensor1}
f\frac{\partial^{2}\Psi_{\rm T}}{\partial r^2}+f'\frac{\partial\Psi_{\rm T}}{\partial r}-\left\{\frac{\ell(\ell+d-3)(\alpha-1)}{(d-4)r^2}+\frac{(\alpha+1)(\alpha-1)}{4r^2}f+\frac{(\alpha+1)}{2r}f'\right\}\Psi_{\rm T}=0~.
\end{equation}
Here, we have suppressed the angular numbers $(\ell,\mathbf{m})$ in the master function. We note that the equation satisfied by the scalar perturbation, presented in \ref{eqn4} and the equation for tensor perturbation, as in \ref{tensor1}, are different for generic $N$, whereas they are the same in the $ d$-dimensional black hole Einstein gravity (for Einstein gravity $N=1$, and $\alpha=d-3$) \cite{Hui:2020xxx}. Thus, it follows that the TLNs will be different for scalar and tensor perturbations. 

Applying a similar technique as in the scalar case, i.e., introducing a rescaled master perturbation function: $u_{\rm T}=x^{-\beta}\Psi_{\rm T}$, where $x\equiv (r_{\rm H}/r)^{\alpha}$ and we arrive at the radial master equation as,
\begin{equation}\label{tensor3}
x(1-x)u_{\rm T}''(x)+\bigg(c-(a+b+1)x\bigg)u_{\rm T}'(x)-abu_{\rm T}(x)=0~,
\end{equation}
where $c\equiv(1/\alpha)+1+2 \beta=a+b$ and, 
\begin{equation}
ab\equiv \frac{\beta}{\alpha}(1+\alpha \beta)+\beta+\frac{\alpha+1}{2\alpha}-\frac{(\alpha+1)(\alpha-1)}{4 \alpha^{2}}\,.
\end{equation}
In general, the coefficient of $u_{\rm T}$ in \ref{tensor3} would depend on $x$, which we have set to zero using the free parameter $\beta$ in order to reduce the differential equation to the form given in \ref{tensor3}. The corresponding equation for $\beta$ reads, 
\begin{equation}
\alpha\beta(1+\alpha\beta)-\frac{\ell(\ell+d-3)(\alpha-1)}{(d-4)}-\frac{(\alpha+1)(\alpha-1)}{4}=0\,,
\end{equation}
which can be solved to obtain the following solution for the parameter $\beta$ as,
\begin{equation}
\beta=-\frac{1}{2\alpha}+\frac{1}{2}(1+2\hat{\ell})\sqrt{1-\frac{\gamma_{2}}{\alpha^2 (1+2 \hat{\ell})^2}}~;\qquad \gamma_{2}\equiv\frac{4 \ell (\ell+1)(d-1)(N-1)}{(d-4)N}\,.
\end{equation}
Here, $\hat{\ell}=(\ell/\alpha)$, which we have used in the context of scalar perturbations as well. Given the above solution for $\beta$, one can determine the constants $a$, $b$ and $c$ in terms of $\alpha$, $\hat{\ell}$ and $\gamma_{2}$, which read,
\begin{equation}
c=2a=2b=1+(1+2\hat{\ell})\sqrt{1-\frac{\gamma_{2}}{\alpha^2 (1+2 \hat{\ell})^2}}
\equiv 2(1+\hat{\ell}+\mathcal{N}_{2})~.
\end{equation}
Here, we have introduced the quantity $\mathcal{N}_{2}$ for notational convenience and has the following expression: 
\begin{equation}
\mathcal{N}_{2}=\frac{1+2\hat{\ell}}{2}\left[\sqrt{1-\frac{\gamma_{2}}{\alpha^2 (1+2 \hat{\ell})^2}}-1 \right]\,.
\end{equation}
In general, the parameters $a$, $b$, and $c$ are not integers for arbitrary values of $N$ and $d$, and in fact they are not even rational numbers, due to the presence of the square root. One possibility for getting rid of the square root and hence possibly obtaining $a$, $b$, and $c$ as integers, is to have $\gamma_{2}=\alpha^{2}(4 \hat{\ell}^{2}+4\hat{\ell})$ which implies that $\ell=3-d$. But our calculation is valid for $d>4$, while $\ell$ is bound to be a positive integer. So $\ell=3-d$ is not a valid choice for our case. However, there exist certain special cases, where these constants $(a,b,c)$ do become integers. We discuss two such cases below.

\begin{itemize}

\item For GR one obtains $N=1$ and hence it follows that:
\begin{equation}
\gamma_{2}=0\,; 
\qquad  
c=2a=2b=\frac{2\ell}{d-3}+2\,.
\end{equation}
Therefore, all the constants $a$, $b$ and $c$ are integers, whenever $\ell/(d-3)$ is also an integer. This result agrees with earlier results in the literature, studied in the context of higher-dimensional Schwarzschild BH \cite{Hui:2020xxx}. As we will show the TLNs in this case vanish identically.

\item A more non-trivial result for pure Lovelock BHs follows from the following relation between the spacetime dimension and the Lovelock order: $d=3N+1$, in which case, one obtains:
\begin{equation}
\gamma_{2}=4\ell(\ell+1)\,; 
\qquad 
c=2a=2b=2\,.
\end{equation}
In this case as well, the parameters $a$, $b$, and $c$ are always integers, irrespective of the Lovelock order $N$. Thus we expect to obtain vanishing TLNs in this case.

\end{itemize}
The situation, where $a$ and $b$ are half-integers, while $c$ is an integer, is something we could not find, as all of them have the square root term. Nonetheless, for completeness we will present the computation of TLNs for all the three cases, i.e., (a) $a$, $b$ and $c$ are all non-integers, (b) $a$ and $b$ are half integers, while $c$ is an integer, and finally (c) $a$, $b$ and $c$ are all integers. We summarize the TLNs for each of these cases below. 
\begin{itemize}

\item \textbf{Case 1:}  For generic values of $a$, $b$, and $c$, the solution of \ref{tensor1} is given in \ref{abcd}. Then for this case the radial part of the tensor perturbation $\Psi_{\rm T}$ can be written as, $\Psi_{\rm T}(r)=(r_{\rm H}/r)^{\alpha\beta}u_{\rm T}(r)$, which has the following asymptotic behaviour (see \ref{response_gen} of \ref{abcd}),
\begin{equation}
\Psi^{\rm (gen)}_{\rm T}(r)=\mathcal{A}(-2\hat{\ell}-1-2\mathcal{N}_{2})\frac{\Gamma(1+2\hat{\ell}+2\mathcal{N}_{2})}{\Gamma(1+\hat{\ell}+\mathcal{N}_{2})^{2}}\left(\frac{r_{\rm H}}{r}\right)^{-\alpha-2\ell-2\alpha\mathcal{N}_{2}+\alpha \beta}\left[1+\frac{\mathcal{F}^{\rm (I)}r_{\rm H}^{\alpha+2\ell+2\alpha\mathcal{N}_{2}}}{r^{\alpha+2\ell+2\alpha\mathcal{N}_{2}}}\right]\,.
\end{equation}
The response function $\mathcal{F}^{\rm (I)}$, is given by \ref{response_I} of \ref{abcd}, from which the TLNs read,
\begin{equation}
k^{\rm tensor\,(gen)}_{\ell}=\frac{1}{2}\textrm{Re}\left(\mathcal{F}^{\rm (I)}\right)=\frac{(2\hat{\ell}+1+2\mathcal{N}_{2})}{2}\frac{\Gamma(1+\hat{\ell}+\mathcal{N}_{2})^{4}}{\Gamma(2+2\hat{\ell}+2\mathcal{N}_{2})^{2}}\frac{\tan[\pi(\hat{\ell}+\mathcal{N}_{2})]}{2\pi}\,.
\label{generallove3}
\end{equation}
Note that the above expression does not work for half-integer values of $\hat{\ell}+\mathcal{N}_{2}$, which we will discuss next. 

\item \textbf{Case 2:} For half-integer values of $a$ and $b$ and integer values of $c$, the asymptotic behaviour of the tensor perturbation becomes,
\begin{eqnarray}
\Psi_{\rm T}^{\rm (mix)}(r)&=&\frac{\pi^{2}B(-1)^{(2\hat{\ell}+2\mathcal{N}_{2})}(1+2\hat{\ell}+2\mathcal{N}_{2})}{\sin^{2}[\pi(\hat{\ell}+\mathcal{N}_{2})]}\times\frac{\Gamma(1+2\hat{\ell}+2\mathcal{N}_{2})^{2}}{\Gamma(1+\hat{\ell}+\mathcal{N}_{2})^{2}}\left(\frac{r_{\rm H}}{r}\right)^{-\alpha-2\ell-2\alpha\mathcal{N}_{2}+\alpha \beta}\nonumber\\&&\times\left[1+\frac{\mathcal{F}^{\rm (II)}r_{\rm H}^{\alpha+2\ell+2\alpha\mathcal{N}_{2}}}{r^{\alpha+2\ell+2\alpha\mathcal{N}_{2}}}\right]\,.
\end{eqnarray}
The above asymptotic expansion for the tensor perturbation yields the response function $\mathcal{F}^{\rm (II)}$, whose real part yields the TLNs associated with the tensor perturbation: 
\begin{equation}
k^{\rm tensor\,(mix)}_{\ell}=(-1)^{2\hat{\ell}+2\mathcal{N}_{2}}\frac{\alpha(2\hat{\ell}+1+2\mathcal{N}_{2})}{2}\frac{\Gamma(1+\hat{\ell}+\mathcal{N}_{2})^{4}}{\Gamma(2+2\hat{\ell}+2\mathcal{N}_{2})^{2}}
\frac{\sin^{2}[\pi(\hat{\ell}+\mathcal{N}_{2})]}{\pi^{2}}\ln\left(\frac{r_{\rm H}}{r}\right)\,.
\end{equation}
We again observe the characteristic Logarithmic piece in the TLNs of this specific scenario. 

\item \textbf{Case 3:} Finally, for integer values of $a$, $b$, and $c$, the radial part of the tensor perturbation, asymptotically has the following behaviour: $\Psi_{\rm T}\sim r^{\alpha+2\ell+2\alpha \mathcal{N}_{2}-\alpha \beta}$, and hence the TLN reads, 
\begin{equation}
k^{\rm tensor\,(int)}_{\ell}=0~.
\end{equation}
Following our previous discussion it is clear that besides BHs in four-dimensional general relativity, the tensorial TLNs of pure Lovelock BHs in $d=3N+1$ vanish identically. 

\end{itemize}

\subsection{Static Love numbers for black holes in pure Lovelock gravity under axial perturbation}

Here, we consider the axial sector of gravitational perturbation, which is also known as vector-type perturbations/odd sector perturbations, for black holes in pure Lovelock theories of gravity. Here also, the quantity $T(r)$ and its derivatives, presented in \ref{T1} and \ref{T2}, play a significant role. Following \cite{Takahashi:2010ye}, it follows that the radial equation for the master function of the axial sector $\Psi_{\rm ax}(r)$ satisfies the following equation in the static case ($\omega=0$),
\begin{eqnarray}
f \frac{\partial^2 \Psi_{\rm ax}}{\partial r^2}+f'\frac{\partial \Psi_{\rm ax}}{\partial r}
-\left\{\frac{\alpha\left[\ell(\ell+d-3)-(d-2)\right]}{(d-3)r^2}+\frac{(\alpha+1)(\alpha+3)}{4r^2}f(r)-\frac{(\alpha+1)}{2r}f'(r)\right\}\Psi_{\rm ax}=0~.
\end{eqnarray}
In terms of a rescaled radial coordinate $x\equiv (r_{\rm H}/r)^{\alpha}$ and a redefined master function: $u_{\rm ax}=x^{-\beta}\Psi_{\rm ax}$, we obtain the following differential equation:
\begin{equation}\label{vector_part}
x(1-x)u_{\rm ax}''(x)+\bigg(c-(a+b+1)x\bigg)u_{\rm ax}'(x)-a b u_{\rm ax}(x)=0~,
\end{equation}
where, $c=(1/\alpha)+1+2 \beta=a+b$, and $ab=(\beta/\alpha)(1+\alpha \beta)+\beta-\{(\alpha+1)/2\alpha\}
-\{(\alpha+1)(\alpha+3)/4\alpha^{2}\}$. In general, the coefficient of $u_{\rm ax}$ will depend on $x$, which will vanish, if and only if $\beta$ satisfies the following algebraic equation:
\begin{equation}
\alpha\beta(1+\alpha\beta)-\frac{\alpha\left[\ell(\ell+d-3)-(d-2)\right]}{(d-3)}-\frac{(\alpha+1)(\alpha+3)}{4} =0\,,
\end{equation}
whose solution yields,
\begin{equation}
\beta= -\frac{1}{2\alpha}+\frac{1}{2}(1+2\hat{\ell})\sqrt{1-\frac{\gamma_{1}}{\alpha^2 (1+2 \hat{\ell})^2}}~;\qquad \gamma_{1}=\frac{4(\ell^2-1)(d-1)(N-1)}{(d-3)N}~.
\end{equation}
Therefore, from the expression for $ab$ and $a+b$ in terms of $(\alpha,\beta)$, we can determine the coefficients $a$, $b$, and $c$ as,
\begin{align}
c&=2+2\hat{\ell}+2\mathcal{N}_{1}=1+(1+2\hat{\ell})\sqrt{1-\frac{\gamma_{1}}{\alpha^2(1+2 \hat{\ell})^2}}~,
\\
a&=2+\hat{\ell}+\mathcal{N}_{1}+\frac{1}{\alpha}=\frac{(1+2 \hat{\ell})}{2}\sqrt{1-\frac{\gamma_{1}}{\alpha^2 (1+2 \hat{\ell})^2}}+\frac{\hat{\ell}}{\ell}+ \frac{3}{2}~,
\\
b&=\hat{\ell}+\mathcal{N}_{1}-\frac{1}{\alpha}=\frac{(1+2 \hat{\ell})}{2}\sqrt{1-\frac{\gamma_{1}}{\alpha^2 (1+2 \hat{\ell})^2}}-\frac{\hat{\ell}}{\ell}-\frac{1}{2}~,
\end{align}
where, we have introduced the quantity $\mathcal{N}_{1}$, in accordance with our discussion in the previous section regarding tensor perturbation, which is defined as, 
\begin{align}
\mathcal{N}_{1}=\frac{1+2\hat{\ell}}{2}\left[\sqrt{1-\frac{\gamma_{1}}{\alpha^2 (1+2 \hat{\ell})^2}}-1 \right]\,.
\end{align}
In general, the parameters $a$, $b$, and $c$ are not integers for arbitrary values of $N$ and $d$. However, there exist certain special cases where they do become integers, e.g., for $N=1$, we get, $\mathcal{N}_{1}=0$, and hence the parameters in \ref{vector_part} become,
\begin{eqnarray}
c=\frac{2\ell}{(d-3)}+2~,
\quad
a=\frac{\ell+1}{(d-3)}+2~,
\quad
b=\frac{\ell-1}{(d-3)}~.
\end{eqnarray}
Therefore, $a$, $b$ and $c$ are integers, when $(\ell+1)/(d-3)$ and $(\ell-1)/(d-3)$ are integers. For $d=3N+1$, unlike the scalar and tensor perturbations, $(a,b,c)$ are not integers for axial perturbation. Rather for $d=2N+3$, we get $\alpha=(2/N)$, $\mathcal{N}_{1}=-\{(\ell-1)(N-1)/2\}$, such that we obtain the parameters in the hypergeometric equation to become,
\begin{eqnarray}
c=\ell+N+1~,
\quad
a=\frac{\ell+3}{2}+N~,
\quad
b=\frac{\ell-1}{2}~.
\end{eqnarray}
Thus, for odd values of $\ell$, the quantities $(a,b,c)$ are integers, while for even choices of $\ell$, $c$ is an integer, but (a,b) are half-integers. We now discuss the cases, when $(a,b,c)$ are non-integers, $c$ is an integer and $(a,b,c)$ are all integers, and determine the TLNs. 

\begin{itemize}
    
\item \textbf{Case 1:}  For generic values of $a$, $b$, and $c$, the solution of \ref{vector_part} and hence the radial part of the master function $\Psi_{\rm ax}$ is given by (see \ref{response_gen} of \ref{abcd})
\begin{align}\label{gen_ax_sol}
\Psi^{\rm (gen)}_{\rm ax}(r)&=\mathcal{A}(-2\hat{\ell}-1-2\mathcal{N}_{1})\frac{\Gamma(1+2\hat{\ell}+2\mathcal{N}_{1})}{\Gamma(2+\hat{\ell}+\mathcal{N}_{1}+\frac{1}{\alpha})\Gamma(\hat{\ell}+\mathcal{N}_{1}-\frac{1}{\alpha})}\left(\frac{r_{\rm H}}{r}\right)^{-\alpha-2\ell-2\alpha\mathcal{N}_{1}+\alpha \beta}
\nonumber
\\
&\qquad \times\left[1+\frac{\mathcal{F}^{\rm (I)}r_{\rm H}^{\alpha+2\ell+2\alpha\mathcal{N}_{1}}}{r^{\alpha+2\ell+2\alpha\mathcal{N}_{1}}}\right]\,,
\end{align}
where, we have used the result $\Psi_{\rm ax}=(r_{\rm H}/r)^{\alpha\beta}u_{\rm ax}(x)$. Given the above asymptotic expansion of the axial master function, one can obtain the response function and hence calculate the TLNs as,
\begin{align}
k^{\rm ax\,(gen)}_{\ell}&=\frac{1}{2}\textrm{Re}\left(\mathcal{F}^{\rm (I)}\right)=\frac{(2\hat{\ell}+1+2\mathcal{N}_{1})}{2}\,\frac{\Gamma(2+\hat{\ell}+\mathcal{N}_{1}+\frac{1}{\alpha})^{2}\,\Gamma(\hat{\ell}+\mathcal{N}_{1}-\frac{1}{\alpha})^{2}}{\Gamma(2+2\hat{\ell}+2\mathcal{N}_{1})^{2}}
\nonumber
\\
&\qquad \times\frac{\sin[\pi(\hat{\ell}+\mathcal{N}_{1}+\frac{1}{\alpha})]\sin[\pi(\hat{\ell}+\mathcal{N}_{1}-\frac{1}{\alpha})]}{\pi\sin[\pi(2\hat{\ell}+2\mathcal{N}_{1})]}\,.
\label{generallove33}
\end{align}
Thus TLNs are non-zero in general. 

\item \textbf{Case 2:} For $a$ and $b$ being half-integers, while $c$ is an integer, the solution of the radial part of axial gravitational perturbation is given by, 
\begin{eqnarray}
\Psi_{\rm ax}^{\rm (mix)}(r)&=&\frac{\pi^{2}B(-1)^{2\hat{\ell}+2\mathcal{N}_{1}}(1+2\hat{\ell}+2\mathcal{N}_{1})}{\sin[\pi(\hat{\ell}+\mathcal{N}_{1}+\frac{1}{\alpha})]\sin[\pi(\hat{\ell}+\mathcal{N}_{1}-\frac{1}{\alpha})]}\times\frac{\Gamma(1+2\hat{\ell}+2\mathcal{N}_{1})^{2}}{\Gamma(2+\hat{\ell}+\mathcal{N}_{1}+\frac{1}{\alpha})\Gamma(\hat{\ell}+\mathcal{N}_{1}-\frac{1}{\alpha})}
\nonumber
\\
&&\times\left(\frac{r_{\rm H}}{r}\right)^{-\alpha-2\ell-2\alpha\mathcal{N}_{1}+\alpha \beta}\left[1+\frac{\mathcal{F}^{\rm (II)}r_{\rm H}^{\alpha+2\ell+2\alpha\mathcal{N}_{1}}}{r^{\alpha+2\ell+2\alpha\mathcal{N}_{1}}}\right]\,.
\end{eqnarray}
From which one can calculate the non-zero axial TLNs as,
\begin{align}\label{axialmix}
k^{\rm ax\,(mix)}_{\ell}&=(-1)^{2\hat{\ell}+2\mathcal{N}_{1}}\,\frac{\alpha(1+2\hat{\ell}+2\mathcal{N}_{1})}{2}\,\frac{\Gamma(2+\hat{\ell}+\mathcal{N}_{1}+\frac{1}{\alpha})^{2}\,\Gamma(\hat{\ell}+\mathcal{N}_{1}-\frac{1}{\alpha})^2}{\Gamma(2+2\hat{\ell}+2\mathcal{N}_{1})^{2}}
\nonumber
\\
&\qquad \times\frac{\sin[\pi(\hat{\ell}+\mathcal{N}_{1}+\frac{1}{\alpha})]\sin[\pi(\hat{\ell}+\mathcal{N}_{1}-\frac{1}{\alpha})]}{\pi^2} \ln \left(\frac{r_{\rm H}}{r}\right)~.
\end{align}

\item \textbf{Case 3:} Finally for integer values of $a$, $b$, and $c$, the solution of \ref{vector_part}, asymptotically has only a growing mode, and hence the TLN vanishes,
\begin{equation}
k^{\rm ax\,(int)}_{\ell}=0~.
\end{equation}

\end{itemize}

The above discusses the TLNs obtained from the axial master function. However, the axial TLNs are typically defined with respect to the perturbation of the $g_{t\phi}$ component of the metric. Here we present the connection between the two. From \cite{Takahashi:2010ye}, we can relate the master function $\Psi_{\rm ax}$ with the metric perturbation $h_{0}$ arising from the perturbation of the $g_{t\phi}$ component, as in the static limit the metric perturbation $h_{1}$ vanishes\footnote{Note that the master function in the static limit need to be scaled by a factor of $(i\omega)$, so that the master function remains non-vanishing in the static limit.}. This yields, 
\begin{eqnarray}\label{ax_conn}
h_{0}&=& \frac{f}{T'}\bigg[\frac{d\Psi_{\rm ax}}{dr}r\sqrt{T'} +\Psi_{\rm ax}\left\{\sqrt{T'}+\frac{rT''}{2\sqrt{T'}}\right\}\bigg]
\nonumber
\\
&=&\frac{f}{\sqrt{T'}}\bigg(\frac{\alpha+1}{2}\Psi_{\rm ax}+ r \frac{d \Psi_{\rm ax}}{d r}\bigg)~,
\end{eqnarray}
where we have used the ratio between $(T'/T)$ as well as $(T''/T')$ in order to arrive at the final expression. In order to find the connection between the Love numbers defined through $\Psi_{\rm ax}$ and $h_{0}$, from \ref{gen_ax_sol} we notice the following asumptotic behaviour of the axial master function: $\Psi_{\rm ax}\sim A_{1} r^{\alpha+2 \ell+2 \alpha \mathcal{N}_{1}-\alpha\beta}+A_{2} r^{-\alpha\beta}$. Substitution of the same in \ref{ax_conn} yields the following asymptotic behaviour for $h_{0}\sim (A_{1}/\sqrt{N\alpha})~[\ell+1+\alpha(\mathcal{N}_{1}+1)]r^{\ell+1+\alpha \mathcal{N}_{1}}+(A_{2}/\sqrt{N\alpha})~[-\ell-\alpha \mathcal{N}_{1}+1] r^{-\ell-\alpha \mathcal{N}_{1}+(1-\alpha)}$. Hence, the TLNs obtained from the master function are related to the TLNs associated with metric perturbations as,
\begin{equation}
k_{\ell}=\bigg(\frac{1-\ell-\alpha \mathcal{N}_{1}}{1+\ell+\alpha (\mathcal{N}_{1}+1)}\bigg)k^{\rm ax}_{\ell},
\end{equation}
where $k^{\rm ax}_{\ell}$ is the tidal Love number obtained from the axial master function. For $d$ dimensional general relativity, we substitute $N=1$, leading to $\mathcal{N}_{1}=0$, and then the master function is given by the Regge-Wheeler function, with the above relation between axial TLNs becomes
\begin{equation}
k_{\ell}=\bigg(\frac{1-\ell}{\ell+d-2 }\bigg)k^{\rm RW}_{\ell}~.
\end{equation}
One can verify that the above result coincides with \cite{Hui:2020xxx}. We now move to the final part, i.e., polar/scalar TLNs for pure Lovelock BHs.

\subsection{Static tidal Love numbers for black holes in pure Lovelock gravity under polar perturbation}

In this final section regarding TLNs of pure Lovelock BHs, we consider the polar sector of gravitational perturbation, which is also known as scalar-type perturbations/even sector perturbations. Here also, the quantity $T(r)$ and its derivatives, presented in \ref{T1} and \ref{T2}, play a significant role. Following \cite{Takahashi:2010ye}, it follows that the radial equation for the master function of the polar sector $\Psi_{\rm pol}(r)$ satisfies the following equation in the static case ($\omega=0$),

\begin{eqnarray}
\bigg(\frac{\partial^2}{\partial r_{*}^2}+\omega^2-V_{\rm s}(r)\bigg)\Psi_{\rm pol}(r)=0~,\label{scalar}
\end{eqnarray}
where the effective potential $V_{\rm s}(r)$ for the scalar-type perturbations can be expressed as,
\begin{eqnarray}
V_s(r)&=&2\ell(\ell+d-3)f\frac{(rKT)'}{(d-2)KTr^2}-\frac{f}{K}\partial_r\left(f\partial_rK \right)+2f^2\frac{K^{'2}}{K^{2}}
\nonumber
\\
&\ &-\frac{f}{T}\partial_r(f\partial_rT)+2f^2\frac{T^{'2}}{T^2}+2f^2\frac{K^{'}T^{'}}{KT}\,,
\end{eqnarray}
where, $T(r)=Nr^{\alpha}$, and the quantity $K(r)$ becomes,
\begin{eqnarray}
K=\frac{A(r)}{r\sqrt{T'}}~;
\qquad
A(r)=-2 (d-2)f+2 \ell (\ell+d-3)+(d-2)rf'~.
\end{eqnarray}
Due to the presence of the quantity $A(r)$ and its derivatives, the differential equation for $\Psi_{\rm pol}$ cannot be expressed in the hypergeometric form. This is why we first write down the differential equation satisfied by the perturbation of the $g_{tt}$ component $\bar{H}$ in the static limit, which takes the following form, 
\begin{eqnarray}
\eta(x)\partial_{x}^{2}\bar{H}(x)+\beta(x)\partial_{x}\bar{H}(x)+\gamma(x)\bar{H}(x)=0~,
\end{eqnarray}
where, we have used the coordinate $x=(r_{\rm H}/r)^{\alpha}$. The coefficients $\eta(x)$, $\beta(x)$ and $\gamma(x)$ can be obtained by using the expressions for $T(r)$, $T'(r)$ and $T''(r)$, along with derivatives of $A(r)$, such that they read, 
\begin{align}
\eta(x)&=\alpha^2(d-2)(x-1)x^2\Big[4(\ell-1)(d+\ell-2)-4x(\ell-1)(d+\ell-2)+x^2(\alpha -1)(d-2)\Big]\,,
\\
\beta(x)&=\alpha^2(d-2)x^2\Big[(\ell-1)(d+\ell-2)+x\big\{2d(\alpha-2\ell+1)-4\big[\alpha+(\ell-3) \ell+1\big]\big\}\Big] \,,
\\
\gamma(x)&=(\alpha +1)\Big[x^2\big\{(3\alpha-1)(d-2)(d-3)\ell-2\alpha^2(d-2)^2+\ell^2(3\alpha-1)(d-2)\big\}
\nonumber
\\
&+(x-1)\big\{4(d-2)(d-3)\ell-4(d^2-7d+11)\ell^2-8(d-3)\ell^3-4\ell^4\big\}\Big]\,.
\end{align}
As evident, in the limit $\alpha \to 1$, and $d\to 4$, we get back the usual result in the context of Schwarzschild BH of general relativity \cite{Silvestrini:2025lbe}. We observe that the differential equation satisfied by $\bar{H}(x)$ is also not in a Hypergeometric form. In order to arrive at a Hypergeometric differential equation, we define the following quantity,
\begin{eqnarray}
\bar{\Psi}_{\rm pol}=F(x)\bar{H}'(x)+G(x)\bar{H}(x)~,
\end{eqnarray}
where $F(x)$ and $G(x)$ are unknown functions of the radial coordinate that need to be determined. For this purpose, we demand that the differential equation for $\bar{\Psi}_{\rm pol}$ must take the following form,
\begin{eqnarray}\label{scalar_part}
x(1-x)\bar{\Psi}_{\rm pol}''(x)+\bigg(c-(a+b+1)x\bigg)\bar{\Psi}_{\rm pol}'(x)-ab\bar{\Psi}_{\rm pol}(x)=0~,
 \end{eqnarray}
where the coefficients $a$, $b$ and $c$ are given by,
\begin{align}
c&=2+2\mathcal{N}_{0}+2\hat{\ell}\,,
\quad
a=\mathcal{N}_{0}+\hat{\ell}+2\,,
\quad
b=\mathcal{N}_{0}+\hat{\ell}\,,
\\
\mathcal{N}_{0}&=\frac{(1+2 \hat{\ell})}{2}\left[\sqrt{1-\frac{\gamma_{0}}{\alpha^2 (1+2 \hat{\ell})^2}}-1\right]\,.
\end{align}
These coefficients are motivated from the tensor ($s=2$) and the axial ($s=1$) perturbations, where we have a quantity $\gamma_{2}$ and $\gamma_{1}$, respectively, whose generalization to the scalar case ($s=0$) is given by, $\gamma_{0}=\{4\ell(\ell-1)(d-1)(N-1)/(d-2)N\}$. Thus in general we have the quantity $\gamma_{s}$ defined as, 
\begin{equation}
\gamma_{s}=\frac{4\big[(\ell+s-1)^{2}-1+(1-s)\ell\big](d-1)(N-1)}{(d-s-2)N}~,
\end{equation}
where, $s=2$ yields the result for the tensor perturbation, $s=1$ for vector perturbation and $s=0$ for the scalar-type gravitational perturbation. 

Given the functions $\eta(x)$, $\beta(x)$ and $\gamma(x)$, along with the constants $a$, $b$ and $c$, we obtain two second order coupled differential equations for the functions $F(x)$ and $G(x)$, which are given by,
\begin{align}
&\eta(x)\Big[\eta(x)\big\{F'(x)\big[c-x(a+b+1)\big]-(x-1)x\big[F''(x)+2G'(x)\big]\big\}+G(x)\big\{\eta(x)\big[c-x (a+b+1)\big]
\nonumber
\\
&\quad +(x-1)x\beta (x)\big\}+2(x-1)x\beta(x) F'(x)\Big]+F(x)\Big[\eta(x)\big\{\beta (x)\big[x(a+b+1)-c\big]
\nonumber
\\
&\quad +(x-1)x\big[\beta'(x)+\gamma(x)\big]\big\}-ab\eta(x)^2-(x-1)x\beta(x)\big\{\eta'(x)+\beta(x)\big\}\Big]=0\,.
\label{fdprime}
\\
&\eta(x)\Big\{\eta(x)\big[G'(x)\big\{x (a+b+1)-c\big\}+(x-1)xG''(x)\big]-2(x-1)x\gamma(x) F'(x)\Big\} 
\nonumber
\\
&\quad +F(x)\Big\{\eta(x)\big[\gamma(x)\big\{c-x (a+b+1)\big\}-(x-1)x\gamma'(x)\big]+(x-1)x\gamma(x)\big[\eta'(x)+\beta(x)\big]\Big\}
\nonumber
\\
&\quad +G(x)\eta(x)\Big\{ab\eta(x)-(x-1)x\gamma(x)\Big\}=0\,.
\label{gdprime}
\end{align}
As evident, \ref{fdprime} involved $F''(x)$ along with first derivatives of $F$ and $G$ and simply $F$ and $G$. While \ref{gdprime} involves $G''$ along with the rest of the terms as in \ref{fdprime}. The validity of \ref{scalar_part} will follow from the existence of well-behaved solutions to \ref{fdprime} and \ref{gdprime}. We have checked that, for generic boundary conditions, which are well-behaved at the horizon as well as asymptotically, the solutions to the above differential equations are also well-behaved throughout the region of interest. This implies that such functions $F(x)$ and $G(x)$ always exist.   

    
    

In general, the parameters $a$, $b$, and $c$ are not integers for arbitrary values of Lovelock order $N$ and spacetime dimension $d$. However, there exist special cases where they do become integers as well as half-integers. First of all, for $N=1$, we get,
\begin{eqnarray}
c=\frac{2 \ell}{(d-3)}+2~,
\quad
a=\frac{\ell}{(d-3)}~,
\quad
b=\frac{\ell}{(d-3)}+2~,
\end{eqnarray}
and hence, $a$, $b$ and $c$ are integers, whenever $\{\ell/(d-3)\}$ is also an integer. In addition for $d=N+1$, the above quantities can be integers. However, non-trivial physics exists for $d\geq 2N+2$ alone, and hence the above spacetime dimension will not give rise to any non-trivial dynamics for gravity. Thus in general, except for the trivial general relativistic limit, the quantities $a$, $b$ and $c$ are never integers. Therefore, in this case we simply present the TLNs for generic $a$, $b$ and $c$. In this case, the solution of \ref{scalar_part} can be obtained from \ref{abcd}, which takes the following form,
\begin{align}
\bar{\Psi}_{\rm pol}^{\rm gen}=\mathcal{A}(-1-2\hat{\ell}-2\mathcal{N}_{0})\frac{\Gamma(1+2\hat{\ell}+2\mathcal{N}_{0})}{\Gamma(\hat{\ell}+\mathcal{N}_{0}+2)\Gamma(\Gamma(\hat{\ell}+\mathcal{N}_{0})}\left(\frac{r_{\rm H}}{r}\right)^{-1-2\hat{\ell}-2\mathcal{N}_{0}}\left[1+\mathcal{F}_{\rm pol}^{\rm gen}x^{1+2\hat{\ell}+2\mathcal{N}_{0}}\right]\,,
\end{align}
where, the response function and hence the TLNs are given by, 
\begin{align}
k_{\ell\,(\textrm{pol})}^{\rm gen}=\frac{1}{2}\mathcal{F}_{\rm pol}^{\rm gen}&=(1+2\hat{\ell}+2\mathcal{N}_{0})\frac{\Gamma(2+\hat{\ell}+\mathcal{N}_{0})^2 \Gamma(\hat{\ell}+\mathcal{N}_{0})^2}{\Gamma(2+2\hat{\ell}+2\mathcal{N}_{0})^2}\frac{\tan(\pi \hat{\ell}+\pi\mathcal{N}_{0})}{4\pi}~.
\end{align}
Note that when $\hat{\ell}+\mathcal{N}_{0}$ is an integer, e.g., in the case of GR with $\ell/(d-3)$ being an integer, it follows that the above TLNs vanish identically. This result is consistent with that of vanishing TLNs for BHs in GR.

To summarize, in general the TLNs of a pure Lovelock BH do not vanish. Unlike the case of GR, where the TLNs generically go to zero whenever the ratio $\ell/(d-3)$ is an integer, for pure Lovelock BHs, even if $\ell/(d-3)$ is an integer, the TLNs do not necessarily vanish. This is because, in the context of pure Lovelock theories, besides the spacetime dimensions $d$ and the angular number $\ell$, the perturbations also depend on the Lovelock order $N$. Thus obtaining zero TLNs under perturbation is much more constrained for pure Lovelock BHs. For example, in the case of the scalar and tensor parts of the gravitational perturbation, the TLNs vanish identically for $d=3N+1$, where $N$ is the order of the Lovelock BH. Similarly, both the scalar and axial parts of gravitational perturbations lead to vanishing TLNs for $d=2N+3$, while the TLNs associated with polar perturbations are non-zero. However, there is not a single relation between the spacetime dimension $d$ and the Lovelock order $N$ for which the TLNs under arbitrary perturbation vanish. This shows that TLNs remain non-zero for a broader parameter space for pure Lovelock theories of gravity. We summarize these findings through \ref{tab:staticLoveLock} providing the cases in which the TLNs of the pure Lovelock BH under scalar, tensor, axial and polar perturbations vanish identically.

\begin{table*} 
\begin{center}  
\begin{tabular}{ccccccccccccccc}        
\hline\hline
\textrm{Perturbation} & \textrm{Dimension} ($d$) &  Lovelock  & Angular  & Static Love number  \\ 
& & order ($N$) & harmonics ($\ell$) & \\ \hline \hline
Scalar & $ 4 $ & $1$ & All values & $0$ \\
          &$3N+1$& All values & All values & $0$\\
          &$2N+2$& All values & All  values & $0$\\
          &$2N+3$&$Even$ & All values & $0$\\
          &$2N+3$&$Odd$ & Even & $0$\\ \hline
$Tensor$ &$3N+1$& All values & All values & $0$\\\hline
$Axial$ & $ 4 $ & $1$ & All values & $0$ \\
          &$2 N+3$& All values & $Odd$ & $0$\\ \hline
$Polar$ & $ 4 $ & $1$ & All values & $0$ \\\hline
          
\end{tabular}                                
\caption{We have shown the specific cases in which the TLNs for scalar, axial, polar and tensor perturbations of a pure Lovelock BH vanish. As mentioned in the text, except for these special cases, the TLNs for the perturbations of the pure Lovelock BHs are non-zero. As evident, in the general relativistic limit, in agreement with previous studies of higher-dimensional Schwarzschild BHs \cite{Hui:2020xxx} the TLNs vanish identically.}
\label{tab:staticLoveLock} 
\end{center}   
\end{table*}

\section{Static tidal Love number for black holes in Einstein-Gauss-Bonnet gravity} \label{Gauss}

In the previous sections we have considered the TLNs of pure Lovelock BHs with arbitrary Lovelock order $N$ and spacetime dimensions $d$. As a generalization of the above setup, which uses a single Lovelock order $N$ in the Lovelock polynomial, in this section, we consider two terms in the Lovelock polynomials, $N=1$ and $N=2$, known as the Einstein-Gauss-Bonnet gravity. For BHs in this theory, we compute the scalar, vector, and tensor modes of gravitational perturbations, as well as scalar perturbations. Based on these perturbations, we determine the static TLNs of static and spherically symmetric BHs in the Einstein-Gauss-Bonnet gravity and assess if these TLNs are zero or not.

The background metric of a static and spherically symmetric BH in the Einstein-Gauss-Bonnet gravity is given by \cite{Guo:2018exx, PhysRevD.65.084014, Cvetic:2001bk, Hendi:2015pda} (see also \ref{App:gravlove}),
\begin{equation}
ds^{2}=-f(r)dt^{2}+f(r)^{-1}dr^2+r^2 d \Omega_{d-2}^{2}~,
\quad
f(r)= 1+ \frac{r^2}{2\tilde{\alpha}}\left[1-\sqrt{1+\frac{64\pi \tilde{\alpha} M}{(d-2)S_{d-2}r^{d-1}}} \right]\,.
\end{equation}
Here, M is the mass of the black hole, $d$ is the spacetime dimension, $S_{d-2}$ is the area of the $(d-2)$ dimensional unit sphere, and $\tilde{\alpha}=(d-3)(d-4)\alpha$, with $\alpha$ being the Gauss-Bonnet coupling constant appearing as the coefficient of the Lovelock Lagrangian at order $N=2$. As evident, in the limit $\tilde{\alpha}\to 0$ the above metric function $f(r)$ reduces to that of a $d$ dimensional Schwarzschild BH. It turns out that, unlike the case of the Schwarzschild and the pure Lovelock BHs, due to the complex structure of the metric element, for the Gauss-Bonnet BH, TLNs can be determined only in the small $\alpha$ limit. In this limit, the metric function $f(r)$ for the Gauss-Bonnet BH takes the following form, 
\begin{equation}
f(r)=1 -\frac{16 \pi  M }{(d-2) \Sigma_{d-2}r^{d-3}}+\tilde{\alpha} \frac{256 \pi ^2  M^2}{(d-2)^2 \Sigma_{d-2} ^2 r^{2 d-4}}=1-\frac{2\mathbb{M}}{r^{d-3}}+\frac{4\mathbb{M}^{2}\tilde{\alpha}}{r^{2d-4}}=1-\left(\frac{r_{\rm s}}{r}\right)^{d-3}+\left(\frac{\tilde{\alpha}}{r_{\rm s}^{2}}\right) \left(\frac{r_{\rm s}}{r}\right)^{2d-4}\,.
\end{equation}
where $r_{s}$ is the radius of the horizon for a black hole in Einstein gravity. We will study perturbations of the Einstein-Gauss-Bonnet BH, as presented above, and shall discuss the associated TLNs. 

\subsection{Love numbers of black holes in Einstein-Gauss-Bonnet gravity for scalar perturbations}

Here, we calculate the static perturbation of the background BH geometry due to an external scalar field and hence determine the static Love number for the Einstein-Gauss-Bonnet BH gravity. The scalar field satisfies the Klein-Gordon equation in curved spacetime, $g^{\mu\nu}\nabla_{\mu}\nabla_{\nu}\Phi=\square\Phi=0$. Given the fact that the background metric does not depend on the time and the azimuthal coordinates, we adopt the following ansatz for the scalar field,
\begin{equation}
\Phi(t,r,\Omega)=\sum_{\ell,m}e^{-i\omega t}\frac{\phi(r)}{r^{(d-2)/2}}\mathcal{Y}_{\ell m}(\Omega)~,
\end{equation}
where, $\mathcal{Y}_{\ell m}(\Omega)$ corresponds to angular harmonics on the $(d-2)$ sphere. Substitution of the above decomposition for the scalar field in the Klein-Gordon equation leads to the following master equation for the radial function $\phi(r)$,
\begin{eqnarray}
&&\left(\frac{\partial^2}{\partial r_{*}^2}+\omega^2-V_{\rm eff}(r)\right)\phi(r)=0\nonumber\\
V_{\rm eff}(r) &=& f(r)\left\{\frac{\ell(\ell+d-3)}{r^2}+\frac{(d-2)(d-4)}{4r^2}f(r)+\frac{(d-2)}{2r}f'(r)\right\}~,
\end{eqnarray}
where $dr_{*}=\{dr/f(r)\}$ is the tortoise coordinate and $V_{\rm eff}(r)$ is the effective potential expressed above in terms of the metric function $f(r)$. Now for the static case ($\omega=0$), we arrive at the following second-order differential equation,
\begin{equation}\label{eqn62}
f \frac{\partial^2 \phi}{\partial r^2}+  f' \frac{\partial\phi}{\partial r}- \left\{\frac{\ell(\ell+d-3)}{r^2}+\frac{(d-2)(d-4)}{4r^2}f(r)+\frac{(d-2)}{2r}f'(r)\right\}\phi=0~.
\end{equation}
Due to the complicated structure of the metric function $f$, we will work with the small $\alpha$ approximation. In which we expand the metric function $f(r)$ as, $f=f^{(0)}+\tilde{\alpha}f^{(1)}$, and also the scalar field as $\phi=\phi_{0}+\tilde{\alpha}\phi_{1}$. Therefore, the above reduced form of the Klein-Gordon equation becomes, at zeroth order in $\widetilde{\alpha}$,
\begin{equation}
f_{0}\frac{\partial^2 \phi_{0}}{\partial r^2}+ f_{0}' \frac{\partial\phi_{0}}{\partial r}- \left\{\frac{\ell(\ell+d-3)}{r^2}+\frac{(d-2)(d-4)}{4r^2}f_{0}+\frac{(d-2)}{2r}f_{0}'\right\}\phi_{0}=0~.
\end{equation}
While at linear order in $\alpha$, the radial part of the Klein-Gordon equation becomes, 
\begin{eqnarray}\label{linalphascalar}
&&f_{0} \frac{\partial^2 \phi_{1}}{\partial r^2}+  f_{0}' \frac{\partial\phi_{1}}{\partial r}- \left\{\frac{\ell(\ell+d-3)}{r^2}+\frac{(d-2)(d-4)}{4r^2}f_{0}(r)+\frac{(d-2)}{2r}f_{0}'(r)\right\}\phi_{1}
\nonumber\\
&&=-\frac{4 \mathbb{M}^{2}}{r^{2 d-4}}\frac{\partial^2 \phi_{0}}{\partial r^2}+  \frac{4 (2 d -4) \mathbb{M}^{2}}{r^{2 d-3}}\frac{\partial\phi_{0}}{\partial r}+\frac{4 \mathbb{M}^{2}}{2 r^{2d-2}}\bigg((2d-4)(d-2)+\frac{(d-2)(d-4)}{2}\bigg)\phi_{0}~,
\end{eqnarray}
where, $f_{0}=1-(r_{\rm s}/r)^{d-3}$ and $f_{1}=(4\mathbb{M}^{2}/r^{2d-4})$. 
The solution of $\phi_{0}$ is well known from earlier results regarding static TLNs of higher-dimensional Schwarzschild BH spacetime, as discussed in \cite{Hui:2020xxx}, which can also be obtained from the previous section with the specific choice of $N=1$. After substitution of $\phi_{0}$ in \ref{linalphascalar}, we obtain the solution for $\phi_{1}$. Once the solution for $\phi_1$ is obtained, multiplying it by $\tilde{\alpha}$ and then adding it to the solution for $\phi_0$ we obtain the full solution for $\phi$. The solution for $\phi_{1}$ is obtained by integrating \ref{linalphascalar} numerically. Subsequently, we impose a regularity condition on the horizon, and then fit the asymptotic solution to extract the coefficients of $r^{-\ell}$ term and $r^{\ell+d-3}$ term. The ratio of the coefficients of $r^{-\ell}$ and $r^{\ell+d-3}$ gives the TLNs. As evident, a general solution to the equation for $\phi_{1}$ cannot be obtained for arbitrary values of $d$ and $\ell$. Thus, we focus on analyzing specific cases:

\begin{itemize}

\item \textbf{Case 1:} For $d=5$ and $\ell=2$, it follows that $\phi_{0}$ satisfies a hypergeometric differential equation, with the corresponding solution, satisfying regularity at the horizon is a purely growing mode. Thus in this case the TLN of five dimensional Schwarzschild BH vanishes. However, in the Einstein-Gauss-Bonnet gravity, due to the presence of the Gauss-Bonnet coupling $\tilde{\alpha}$, the TLN becomes non-zero with a logarithmic contribution and is given by
\begin{equation}
k^{\rm EGB}_{2} = \frac{\tilde{\alpha}\left[528 \log \left(\frac{r_{\rm s}}{r} \right) + 125 \right]}{144r_{\rm s}^2}~.
\end{equation}
    
    
\item \textbf{Case 2:} For $d=6$ and $\ell=2$, again the solution of $\phi_{0}$ is a hypergeometric function, which has both growing as well as decaying modes, asymptotically. Thus, the six-dimensional Schwarzschild BH has a non-zero tidal Love number and is given by $k^{\rm EH}_{2}=0.28$. In the Einstein-Gauss-Bonnet gravity, it receives a correction proportional to $\tilde{\alpha}$ and becomes,
\begin{equation}
k^{\rm EGB}_{2} = 0.28 + \tilde{\alpha}~0.12~,
\end{equation}
where we have used the result that $r_{\rm s}^{d-3}=2\mathbb{M}$ and have taken the mass $\mathbb{M}$ to be unity.   

\item \textbf{Case 3:} For $d = 7$ and $\ell = 2$, the TLN of the higher dimensional Schwarzschild BH, obtain from the hypergeometric solution of $\phi_{0}$ is given by $k_{2}^{\rm EH}=0.5\log(r_{0}/r)$. Due to the corrections from the Gauss-Bonnet parameter, in Einstein-Gauss-Bonnet gravity, it is modified to
\begin{equation}
k^{\rm EGB}_{2}=0.5\log\left(\frac{r_{0}}{r}\right)+\tilde{\alpha}\times0.26\times \log\left(\frac{r_{0}}{r}\right)~,
\end{equation}
with $r_{\rm s}^{d-3}=2\mathbb{M}$ and $r_{0}$ being an arbitrary scale associated with the radial perturbation equation.

\end{itemize}

The above depicts that in general, irrespective of the spacetime dimensions $d$ and the angular number $\ell$, the scalar TLNs of Einstein-Gauss-Bonnet BH are non-zero. We would like to verify next, if this holds true for gravitational perturbations as well. 

\subsection{Love numbers of black holes in Einstein-Gauss-Bonnet gravity under tensor perturbation}

Here, we consider the tensor part of the gravitational perturbation of BHs in Einstein-Gauss-Bonnet gravity. The equation governing the tensor perturbation is given by \cite{Takahashi:2010ye},  
\begin{eqnarray}\label{tensor2}
\bigg(\frac{\partial^2}{\partial r_{*}^2}+\omega^2-V_{\rm t}(r)\bigg)\Psi_{\rm T}(r)=0~;
\quad
V_t(r)=\frac{\ell(\ell+d-3)f}{(d-4)r}\frac{d \ln{T^{'}}}{dr}+\frac{1}{r\sqrt{T^{'}}}f\frac{d}{dr}\left(f\frac{d}{dr}r\sqrt{T^{'}}\right) \ .
\end{eqnarray}
where, $V_{t}$ is the effective potential and the quantity $T$ and its derivatives for a small value of $\tilde{\alpha}$ are given by,
\begin{eqnarray}\label{TEGB}
T(r)&=&\frac{ (d-2) r^{d-3}}{2}\left(1+\frac{2 \tilde{\alpha} r_{\rm s}^{d-3}}{r^{d-1}}\right)\,;
\quad
\frac{T'}{T}=\frac{d-3}{r}-2 \tilde{\alpha}  (d-1) r^{-d}r_{\rm s}^{d-3}\,,
\nonumber\\
\frac{T''}{T'}&=&\frac{d-4}{r}+\frac{4 \tilde{\alpha}  (d-1) r^{-d}r_{\rm s}^{d-3}}{d-3}\,.
\end{eqnarray}
Using the above results for $T(r)$, $T'(r)$ and $T''(r)$, we get the radial master equation for the tenosr part of the gravitational perturbation to read, 
\begin{eqnarray}
f \frac{\partial^2 \Psi_{\rm T}}{\partial r^2}&+&  f' \frac{\partial\Psi_{\rm T}}{\partial r}- \bigg\{\frac{(d-2)(d-4) f(r)+2 (d-2) r f'(r)+4 l (d+l-3)}{4 r^2}\nonumber\\&+&\frac{2 \alpha  (d-1) r^{-d-1} r_{\rm s}^{d-3}\left(d r f'(r)-2 d f(r)+2 d \ell-4 r f'(r)+8 f(r)+2 \ell^2-6 \ell\right)}{(d-4) (d-3)}\bigg\}\Psi_{\rm T}=0~.\nonumber\\
\end{eqnarray}
In order to determine the TLNs arising out of the above perturbation equation, we adopt an approach similar to that used in the scalar case. The zeroth order in $\tilde{\alpha}$ term, $\Psi_{\rm T}^{(0)}$ satisfies a hypergeometric equation, which is identical to \ref{tensor1}, with $N=1$. However, the equation at linear order in the Gauss-Bonnet coupling constant can be expressed as,
\begin{align}
f_{0}&\frac{\partial^2 \Psi^{(1)}_{\rm T}}{\partial r^2}+f_{0}'\frac{\partial \Psi^{(1)}_{\rm T}}{\partial r}-\frac{(d-2)(d-4) f_0+2 (d-2) r f_0'+4 l (d+l-3)}{4 r^2}\Psi^{(1)}_{\rm T}
\nonumber\\
&=-\frac{r_{\rm s}^{d-3}}{8r^{2 d+1}}\left(\frac{32 (d-1) \{d (l-1)+(l-3) l+4\}r^d}{(d-4)(d-3)}
+\frac{16(d-1)^2 r^{d+3} r_{\rm s}^{d-3}}{(d-3)r^d}-(d-2)(3d-4) r^3 r_{\rm s}^{d-3}\right)\Psi^{(0)}_{\rm T}
\nonumber\\
&+2(d-2)r^{3-2 d} r_{\rm s}^{2 d-6} \frac{\partial\Psi^{(0)}_{\rm T}}{\partial r}-r^{4-2 d} r_{\rm s}^{2 d-6}\frac{\partial^2 \Psi^{(0)}_{\rm T}}{\partial r^2}~.
\end{align}
In contrast to the pure Lovelock theories, we observe that for BH in Einstein-Gauss-Bonnet theory, the scalar and the tensor perturbations differ, which implies that the corresponding TLNs will also differ. Once the solution for $\Psi^{(1)}_{\rm T}$ is determined, the complete solution for $\Psi_{\rm T}$ can be constructed by multiplying $\Psi^{(1)}_{\rm T}$ with the Gauss-Bonnet coupling $\tilde{\alpha}$ and adding it to the Einstein gravity solution $\Psi^{(0)}_{\rm T}$. As in the previous case, a general solution to the above equation cannot be obtained for arbitrary values of $d$ and $\ell$. Thus we follow the numerical method as detailed in the scalar case to determine the TLNs of Einstein-Gauss-Bonnet BH.  The tidal Love number can then be extracted from the full solution. Consequently, we focus on analyzing specific cases:

\begin{itemize}

\item \textbf{Case 1:} For $d = 5$ and $\ell = 2$, the TLN vanishes in Einstein gravity. However, in Einstein-Gauss-Bonnet gravity, due to the presence of the Gauss-Bonnet coupling $\tilde{\alpha}$, the tidal Love number becomes non-zero and is given by
\begin{equation}
k^{\rm EGB}_{2} = \frac{\tilde{\alpha} \left[1-12\log \left(\frac{r}{r_{\rm s}}\right) \right]}{18 r_{\rm s}^2}~.
\end{equation}

    
\item \textbf{Case 2:} For $d=6$ and $\ell=2$, the zeroth order solution $\Psi_{\rm T}$ involves both ingoing and outgoing solution. So, the tidal Love number is non-zero even in Einstein gravity, with a value of $k_{2}^{\rm EH}=0.28$. In Einstein-Gauss-Bonnet gravity, it receives a linear correction in $\tilde{\alpha}$, resulting in
\begin{equation}
k^{\rm EGB}_{2}= 0.28 +0.14\, \tilde{\alpha}~,
\end{equation}
where, we set $r_{\rm s}^{d-3}=2\mathbb{M}$, where $\mathbb{M}$ is the rescaled mass of the BH, is set to unity.

\item \textbf{Case 3:} For $d=7$ and $\ell=2$, the TLN in Einstein gravity behaves logarithmically, $k_{2}^{\rm EH}=0.5\log(r_{0}/r)$. In Einstein-Gauss-Bonnet gravity, on the other hand, this expression is modified to
\begin{equation}
k^{\rm EGB}_{2}=0.5\log\left(\frac{r_{0}}{r}\right)+\tilde{\alpha}\times0.45\times \log\left(\frac{r_{0}}{r}\right)~,
\end{equation}
with $r_{\rm s}^{d-3}=2\mathbb{M}$, being the horizon of the Schwarzschild BH and $r_{0}$ is an arbitrary characteristic radius.

\end{itemize}

As evident, in all of these cases, the TLNs of BHs in Einstein-Gauss-Bonnet theory, under the tensor part of the gravitational perturbation differ from the TLNs under the scalar perturbation.

\subsection{Love numbers of black holes in Einstein-Gauss-Bonnet gravity under axial perturbation}

Here, we consider the vector-type perturbations (odd sector) for gravitational perturbation of BHs in Einstein-Gauss-Bonnet gravity \cite{Takahashi:2010ye}. The master equations describing axial gravitational perturbation take the following form
\begin{equation}\label{odd}
\bigg(\frac{\partial^2}{\partial r_{*}^2}+\omega^2-V_{\rm ax}(r)\bigg)\Psi_{\rm ax}(r)=0~, 
\quad 
V_{\rm ax}(r)=r\sqrt{T^{'}}f\partial_r \left(f\partial_r\frac{1}{r\sqrt{T^{'}}}\right)+\left(\frac{\ell(\ell+d-3)-1}{d-3}-1\right)\frac{fT^{'}}{rT}\,,
\end{equation}
where $V_{\rm ax}$ is the effective potential for the axial sector of gravitational perturbations. Using the small $\alpha$ expansion of $T(r)$, and its derivatives from \ref{TEGB}, we arrive at the following structure of \ref{odd}, given below,
\begin{eqnarray}
f\frac{\partial^2 \Psi_{\rm ax}}{\partial r^2}&+&  f' \frac{\partial \Psi_{\rm ax}}{\partial r}- \bigg\{\frac{-2 (d-2) r f'(r)+d (d-2) f(r)+4 (\ell-1) (d+\ell-2)}{4 r^2}\nonumber\\&-&\frac{2 \alpha  (d-1) r^{-d-1} r_{s}^{d-3} \left(-2 (d-1) f(r)+(\ell-1) (d+\ell-2)+r f'(r)\right)}{d-3}\bigg\}\Psi_{\rm ax}=0~.
\end{eqnarray}
Here also we will apply a technique similar to the scalar case. The zeroth order equation, will be identical to that of higher dimensional Schwarzschild \cite{Hui:2020xxx}, while we write down the linear-in-$\alpha$ equation as,
\begin{eqnarray}
&&f_{0}\frac{\partial^2 \Psi^{(1)}_{\rm ax}}{\partial r^2}+f_{0}' \frac{\partial\Psi^{(1)}_{\rm ax}}{\partial r}
-\frac{-2 (d-2) r f_{0}'+d (d-2) f_{0}+4 (\ell-1) (d+\ell-2)}{4 r^2}\Psi^{(1)}_{\rm ax}
\nonumber\\
&=&-\bigg[\frac{r^{-2d-1}r_{\rm s}^{d-9}}{8 (d-3)}\bigg\{16 (d-1)r_{\rm s}^6 [d (\ell-3)+\ell (\ell-3)+4] r^d+16 (d-1) (3 d-5) r_{\rm s}^3 r^{d+3} \left(\frac{r_{\rm s}}{r}\right)^d
\nonumber\\
&-&(d-3) (d-2) (5 d-8) r^3 r_{\rm s}^{d+3}\bigg\}\Psi^{(0)}_{\rm ax}-2(d-2)r^{3-2 d}r_{\rm s}^{2d-6} \frac{\partial \Psi^{(0)}_{\rm ax}}{\partial r}+r^{4-2d} r_{\rm s}^{2d-6}\frac{\partial^2 \Psi^{(0)}_{\rm ax}}{\partial r^2}\bigg]~.
\end{eqnarray}
Thus having solved for $\Psi^{(0)}_{\rm ax}$, one can substitute it in the above equation and hence solve it to determine $\Psi^{(1)}_{\rm ax}$. The complete solution for $\Psi_{\rm ax}$ can be constructed by multiplying $\Psi^{(1)}_{\rm ax}$ with the Gauss-Bonnet coupling $\tilde{\alpha}$ and adding it to the Einstein gravity solution $\Psi^{(0)}_{\rm ax}$. The TLNs can then be extracted from the full solution. A general solution for $\Psi^{(1)}_{\rm ax}$ cannot be obtained for arbitrary values of $d$ and $\ell$, consequently, we focus on analyzing specific cases:

\begin{itemize}

\item \textbf{Case 1:} For $d = 5$ and $\ell = 2$, the TLN in Einstein gravity is given by $k_{2}^{\rm EH}=4.69 \log\left(r_{0}/r\right)$. In Einstein-Gauss-Bonnet gravity, it receives a correction proportional to $\tilde{\alpha}$ and becomes
\begin{equation}
k^{\rm EGB}_{2} = 4.69 \log\left(\frac{r_{0}}{r}\right) + 3\, \tilde{\alpha} \log\left(\frac{r_{0}}{r}\right)~,
\end{equation}
with $r_{\rm s}^{d-3}=2\mathbb{M}$, being the horizon of the Schwarzschild BH. A similar logarithmic running behavior is also observed for $d = 7$, $\ell = 2$ and $d = 8$, $\ell = 5$, which holds even within Einstein gravity.

\item \textbf{Case 2:} For $d = 5$ and $\ell = 3$, the solution for $\Psi^{(0)}_{\rm ax}$ is purely growing and hence the TLN vanishes in Einstein gravity. Intriguingly, it also remains zero in the Einstein-Gauss-Bonnet gravity.

\end{itemize}

These results involving TLNs are obtained by using numerical techniques. We put the regularity condition on the horizon and solve for $\Psi^{(1)}_{\rm ax}$. In the asymptotic large-$r$ regime, the numerical solution is expanded and fitted to the power-law forms $r^{-\ell}$ and $r^{\ell + d - 3}$ to determine the associated coefficients, and the ratio of the coefficient of $r^{-\ell}$ to the coefficient of $r^{\ell + d - 3}$ yields the Love number.

\subsection{Love numbers of black holes in Einstein-Gauss-Bonnet gravity under polar perturbation}

In this section, we explore the TLNs associated with the scalar-type perturbations (even sector) for BHs in the Einstein-Gauss-Bonnet gravity. The relevant master equation is as follows \cite{Takahashi:2010ye},
\begin{eqnarray}\label{scalar1}
\bigg(\frac{\partial^2}{\partial r_{*}^2}+\omega^2-V_{\rm s}(r)\bigg)\Psi_{\rm pol}(r)=0~,
\quad
V_{\rm s}(r)&=&2(l(l+d-3))f\frac{(rKT)^{'}}{(d-2)KTr^2}-\frac{f}{K}\partial_r\left(f\partial_rK \right)
\nonumber\\
&+&2f^2\frac{K^{'2}}{K^{2}}-\frac{f}{T}\partial_r(f\partial_rT)
+2f^2\frac{T^{'2}}{T^2}+2f^2\frac{K^{'}T^{'}}{KT}\,.
\end{eqnarray}
where $V_{\rm s}$ is the effective potential for the polar sector of gravitational perturbations. Here, the quantity $T$ and its derivatives are given by \ref{TEGB}, with $K=(A/r\sqrt{T'})$, where, $A=(d-2)(rf'-2f)+2\ell(\ell+d-3)$. Using these results, \ref{scalar1} can be rewritten as,
\begin{eqnarray}
f \frac{\partial^2 \Psi_{\rm pol}}{\partial r^2}&+&  f' \frac{\partial\Psi_{\rm pol}}{\partial r}- \bigg\{\frac{ \left[-2 (d-4) r f'+(d-4) (d-2) f+4 \ell (d+\ell-3)\right]}{4 r^2}
\nonumber\\
&-&\frac{2\tilde{\alpha}  (d-1) r^{-d-1}r_{\rm s}^{d-3} \left[-(d-2) r f'(r)+2 (d-2)^2 f(r)+2 \ell (d+\ell-3)\right]}{d-3}\bigg\}\Psi_{\rm pol}=0~.
\end{eqnarray}
Expanding both $f$ and $\Psi_{\rm pol}$ in powers of $\tilde{\alpha}$, we obtain a zeroth-order equation, which is identical to that of Schwarzschild BH \cite{Hui:2020xxx}, and a first-order equation, which is given by,
\begin{eqnarray}
&&f_{0} \frac{\partial^2 \Psi^{(1)}_{\rm pol}}{\partial r^2}+  f_1' \frac{\partial \Psi^{(1)}_{\rm pol}}{\partial r}-\frac{ \left[-2 (d-4) r f_{1}'+(d-4) (d-2) f{1}+4 \ell (d+\ell-3)\right]}{4 r^2}\Psi^{(1)}_{\rm pol}
\nonumber\\
&=&-\bigg[\frac{r^{-2 d-1}r_{s}^{d-9}}{8 (d-3)} \bigg\{32 (d-1) r_{s}^6 \bigg[(d-3) \ell+(d-2)^2+\ell^2\bigg] r^d-16 (d-2) (d-1) (3 d-7) r_{s}^3 r^{d+3} \bigg(\frac{r_{s}}{r}\bigg)^d\nonumber\\&-&5 (d-4) (d-3) (d-2) r^3 r_{s}^{ d+3}\bigg\}\Psi^{(0)}_{\rm pol}-2 (d-2) r^{3-2 d} r_{s}^{2 d-6} \frac{\partial \Psi^{(0)}_{\rm pol}}{\partial r}+r^{4-2 d} r_{s}^{2 d-6}\frac{\partial^2 \Psi^{(0)}_{\rm pol}}{\partial r^2}\bigg]~.
\end{eqnarray}
An analytical solution to $\Psi^{(1)}_{\rm pol}$ cannot be obtained, due to the complicated source term. Thus we solve it numerically for specific choices of $d$ and $\ell$. The asymptotic structure of the numerical solution is matched with an appropriate growing and decaying modes asymptotically, yielding the TLN. The results of our analysis are as follows:

\begin{itemize}
    
\item \textbf{Case 1:} For $d = 5$ and $\ell = 2$, the polar TLN vanishes in Einstein gravity. However, in Einstein-Gauss-Bonnet gravity, due to the presence of the Gauss-Bonnet coupling $\tilde{\alpha}$, the corresponding TLN becomes non-zero and is given by,
\begin{equation}
k^{\rm EGB}_{2}=\frac{4 \tilde{\alpha} \left[6 \log \left(\frac{r_{\rm s}}{r} \right) - 1 \right]}{9 r_{\rm s}^2}~.
\end{equation}

    
\item \textbf{Case 2:} For $d = 6$ and $\ell = 2$, the polar TLN is non-zero in Einstein gravity and is given by $k^{\rm EH}_{2} =1.74$. In Einstein-Gauss-Bonnet gravity, it receives a correction proportional to $\tilde{\alpha}$ and becomes
\begin{equation}
k^{\rm EGB}_{2}=1.74+\tilde{\alpha}~0.53 ~,
\end{equation}
where we have taken $r_{\rm s}^{d-3}=2$ and the mass $\mathbb{M}$ of the black hole is set to unity.
    
\item \textbf{Case 3:} For $d = 7$ and $\ell = 2$,  the TLN in Einstein gravity is again non-zero and given by, $k_{2}^{\rm EH}=4.5 \log\left(r_0/r\right)$. In Einstein-Gauss-Bonnet gravity, it is modified to
\begin{equation}
k^{\rm EGB}_{2}=4.5\log\left(\frac{r_0}{r}\right) + \tilde{\alpha}\times1.45 \log\left(\frac{r_0}{r}\right)~,
\end{equation}
with $r_{\rm s}^{d-3}=2$, since for this analysis, we have set $\mathbb{M}=1$.

\end{itemize}

To summarize, in general, the TLNs of Einstein-Gauss-Bonnet BH under various perturbations are non-zero, even when the corresponding perturbation of the Einstein BH vanishes, except for some special cases associated with axial perturbations of the Einstein-Gauss-Bonnet BH.  Intriguingly, for \emph{axial} perturbations of the EGB BH if the angular momentum and the spacetime dimension satisfy the following condition $\{(\ell-1)/(d-3)\}=\textrm{integer}$, the TLNs vanish identically. These are the only cases in which the TLNs of an EGB BH vanish alike their Einstein counterpart.  We summarize our findings in \ref{tab:staticEGB}.

\begin{table*} 
\begin{center} 
\begin{tabular}{ccccccccccccccc}        
\hline\hline
Perturbation & Dimension ($d$)&  Angular & Static TLNs & Static TLNs \\ 
& & harmonics ($\ell$) & Einstein Gravity & Einstein-Gauss-Bonnet Gravity \\
\hline
Scalar & $ 5 $&  $2$ & $0$&  $\frac{\tilde{\alpha}\left[528 \log \left(\frac{r_{\rm s}}{r} \right) + 125 \right]}{144r_{\rm s}^2}$ \\
          &$6$& $2$&  $0.28$&  $ 0.28 + \tilde{\alpha}~0.12$\\
          &$7$&$2$ &$0.5\log(r_{0}/r)$ &  $0.5\log\left(\frac{r_{0}}{r}\right)+\tilde{\alpha}\times0.26\times \log\left(\frac{r_{0}}{r}\right)$\\ \hline
Tensor &$ 5 $ & $2$ & $0$ & $\frac{\tilde{\alpha} \left[1-12\log \left(\frac{r}{r_{\rm s}}\right) \right]}{18 r_{\rm s}^2}$ \\
          &$6$&$2$ & $0.28$ & $ 0.28 + \tilde{\alpha}~0.14$\\
         &$7$&$2$ & $0.5\log(r_{0}/r)$ & $0.5\log\left(\frac{r_{0}}{r}\right)+\tilde{\alpha}\times0.45\times \log\left(\frac{r_{0}}{r}\right)$\\ \hline
Axial &$ 5 $ & $2$ & $4.69 \log\left(r_{0}/r\right)$ & $ 4.69 \log\left(\frac{r_{0}}{r}\right) + 3\, \tilde{\alpha} \log\left(\frac{r_{0}}{r}\right)$ \\
          &\textbf{5}&\textbf{3} & \textbf{0} & \textbf{0} \\ \hline
Polar&$ 5 $ & $2$ & $0$ & $\frac{4 \tilde{\alpha} \left[6 \log \left(\frac{r_{\rm s}}{r} \right) - 1 \right]}{9 r_{\rm s}^2}$ \\
          &$6$&$2$ & $1.74$ & $ 1.74 + \tilde{\alpha}~0.53$\\
         &$7$&$2$ & $4.5\log(r_{0}/r)$ & $4.5\log\left(\frac{r_{0}}{r}\right)+\tilde{\alpha}\times1.45\times \log\left(\frac{r_{0}}{r}\right)$\\ \hline
          
\end{tabular}                                
\caption{The above depicts the TLNs for all possible perturbations of the EGB BH. As evident, the TLNs are in general non-zero, except for the cases satisfying $\{(\ell-1)/(d-3)\}=\textrm{integer}$ in the axial perturbations.}
\label{tab:staticEGB}    
\end{center}       
\end{table*}               
     

\section{Discussion} \label{conclusion}

Numerous investigations within the framework of classical GR have demonstrated that the static TLNs of BHs in asymptotically flat, four-dimensional spacetime, are identically zero. This result, which initially emerged from the analyses involving Schwarzschild BH, subjected to external static tidal perturbation, reveals a rather surprising property for BHs in four-dimensional GR, namely, the gravitational potential of these BHS remains monopolar under tidal perturbations, at least in the static regime. This feature has attracted considerable interest in recent years, particularly due to its potential relevance in gravitational wave physics, where TLNs play a crucial role in the late inspiral dynamics of compact binary systems. The vanishing of the TLNs for four-dimensional GR BHs has even been suggested as a potential discriminator between BHs and other compact objects, such as neutron stars or some exotic alternatives, e.g., gravastar, which generally possess nonzero tidal deformation.

However, this vanishing of TLN for BHs appears to be highly sensitive to the dimensionality of the spacetime and the specific gravitational theory under consideration. For example, BHs in GR when extended to higher spatial dimensions, lead to non-zero TLNs. At the same footing, studies have shown that static BHs in theories beyond GR typically possess nonzero TLNs. Both of these features indicate a rich structure for the tidal response of a generic BH compared to its four-dimensional GR counterparts. 

In this work, we have extended the investigation of static TLNs for BHs in GR, to BHs in higher curvature, ghost-free, theories of gravity. These theories of gravity are known as Lovelock theories and we have mainly focused on two broad classes of higher-curvature gravity theories---(a) pure Lovelock gravity and (b) EGB gravity. These theories, as emphasized before, lead to second-order gravitational field equations and avoid ghosts. 
Our analysis focuses on the tidal deformation of static and spherically symmetric BH solutions in these theories. We systematically compute the TLNs by examining the asymptotic behavior of metric perturbations in response to external tidal fields. Remarkably, we find that in certain cases, specifically for particular values of the Lovelock order and spacetime dimension, the static TLNs vanish for BHs in pure Lovelock gravity. This result generalizes the well-known vanishing behavior for BHs in four-dimensional general relativity and suggests that the absence of tidal deformation may persist in more complicated theories of gravity as well under the right conditions. However, unlike GR, where both the axial and the polar TLNs vanish, for pure Lovelock BHs, the TLNs for a specific choice of $d$ and $N$, vanish for axial perturbations but never for polar perturbations. This is a striking feature that can distinguish between BHs in GR to BHs in Lovelock theories of gravity.



In a contrasting manner, our results indicate that within Einstein–Gauss–Bonnet (EGB) gravity, the TLNs are generically non-vanishing, except for a few finely tuned configurations, i.e., when $\{(\ell-1)/(d-3)\}=\textrm{integer}$ within the axial perturbations. By contrast, in pure Lovelock gravity, the TLNs vanish in several sectors, including scalar, tensor and axial perturbations for specific relations between the spacetime dimension, Lovelock order and the angular harmonics. While, in general, the TLNs of a pure Lovelock BHs are also non-zero.

This difference highlights a key distinction between the two theories: while pure Lovelock gravity may, under certain circumstances and specific perturbations, inherit the ``tidally rigid" nature of four-dimensional BHs, Einstein-Gauss-Bonnet gravity does not share this trait. The non-vanishing TLNs in EGB gravity suggest that BHs in this theory always exhibit a finite response to external tidal perturbations, thereby modifying their gravitational signatures in scenarios such as BH binaries.

These findings underscore the sensitivity of TLNs to the underlying gravitational dynamics and may offer new avenues for probing higher curvature corrections to GR through observations of gravitational waves. In particular, precise measurements of tidal effects in BH mergers could, in principle, help distinguish between different theories of gravity in the strong field regime. Our results contribute to the growing body of work seeking to understand the fundamental gravitational properties of compact objects beyond GR, and they motivate further studies of tidal interactions in broader classes of gravitational theories, e.g., studying dynamical tides, or, tides of exotic compact objects in theories beyond GR, along with effect of modified theories on TLN of neutron stars and their universality. We leave these for the future.


\section*{Acknowledgements}
The research of SC is supported by MATRICS (MTR/2023/000049) and Core Research Grants\\
(CRG/2023/000934) from SERB, ANRF, Government
of India. We would like to thank Naresh\\ Dadhich for helpful discussions.

\appendix
\labelformat{section}{Appendix #1} 
\labelformat{subsection}{Appendix #1}
\section{Solution of the hypergeometric equation and its asymptotic behaviour}\label{abcd}

As we have seen in the main text, the determination of static TLNs is intimately connected with the hypergeometric differential equation and the properties of its solution. A hypergeometric differential equation takes the form $x(1-x)u''(x)+[c-(1+a+b)x]u'(x)-abu(x)=0$, where, the properties of the solution are determined by the three constants $(a,b,c)$. For generic choices of these constants, the general solution to the hypergeometric differential equation is given by [The Digital Library of Mathematical Functions (DLMF) (15.10.2) and (15.10.3) \cite{NIST:DLMF}],
\begin{align}
\def\arraystretch{1.2}
u^{\rm (I)}(x)&=A\, {}_2F_1\left(a,b;c;x\right)+B\,x^{1-c}\,{}_2F_1\left(a-c+1,b-c+1;2-c;x\right)
\nonumber
\\
&=A\,{}_2F_1\left(a,b;c;x\right)+B\,x^{1-c}\,{}_2F_1\left(1-b,1-a;2-a-b;x\right)~.
\label{hypergcase1}
\end{align}
In arriving at the second line we have used the fact that, for our case, $c=a+b$. Given the above form for the hypergeometric functions, one can expand both of them about $x=1$ (in our case, this is the horizon), as follows [DLMF (15.8.10) \cite{NIST:DLMF}]:
\begin{align}
{}_2F_1\left(a,b;a+b;x\right)&=-\frac{\Gamma(a+b)}{\Gamma(a)\Gamma(b)} \sum^{\infty}_{k=0}
\frac{\Gamma(a+k)}{\Gamma(a)}\frac{\Gamma(b+k)}{\Gamma(b)} \frac{(1-x)^k}{(k!)^2} 
\nonumber
\\
&\times \left[ \ln(1-x) - 2 \psi(1+k) + \psi(a+k) + \psi(b+k) \right]\,, 
\\
{}_2F_1\big(1-b,1-a;&2-a-b;x\big)=-\frac{\Gamma(2-a-b)}{\Gamma(1-a)\Gamma(1-b)} \sum^{\infty}_{k=0}
\frac{\Gamma(1-a+k)}{\Gamma(1-a)} \frac{\Gamma(1-b+k)}{\Gamma(1-b)} \frac{(1-x)^k}{(k!)^2} 
\nonumber
\\
&\qquad \qquad \qquad \times \left[\ln(1-x) - 2 \psi(1+k) + \psi(1-a+k) + \psi(1-b+k) \right]~.
\end{align}
Note that, only the $k=0$ term is ill-behaved in the $x \to 1$ limit, due to the $\ln(1-x)$ term. While for $k>0$, the term $(1-x)^{k}\ln(1-x)$ vanishes in the $x \to 1$ limit. Thus the solution $u(x)$ in the $x\to1$ limit reads, 
\begin{equation}
u^{\rm (I)}(x\to 1)=-A\frac{\Gamma(a+b)}{\Gamma(a)\Gamma(b)}\ln(1-x)-B\frac{\Gamma(2-a-b)}{\Gamma(1-a)\Gamma(1-b)}\ln(1-x)+\textrm{finite~terms}\,.
\end{equation}
The regularity of $u^{\rm (I)}(x)$ in the $x\to 1$ limit, i.e., at the horizon, demands the following relation between the coefficients:
\begin{equation}
A=\mathcal{A}\frac{\Gamma(2-a-b)}{\Gamma(1-a)\Gamma(1-b)}~; 
\quad
B=-\mathcal{A}\frac{\Gamma(a+b)}{\Gamma(a)\Gamma(b)}~.
\end{equation}
Therefore, the solution $u^{\rm (I)}(x)$, which is regular at the horizon, reads,
\begin{align}
u^{\rm (I)}(x)&=\mathcal{A}\bigg[\frac{\Gamma(2-a-b)}{\Gamma(1-a)\Gamma(1-b)}\,{}_2F_1\left(a,b;c;x\right)
-\frac{\Gamma(a+b)}{\Gamma(a)\Gamma(b)}\,x^{1-c}\,{}_2F_1\left(1-b,1-a;2-c;x\right)\bigg]~.
\label{hypergcase2}
\end{align}
Again, we have the following identity [DLMF (15.10.17)\cite{NIST:DLMF}]
\begin{align}
{}_2F_1\left(a,b;1;1-x\right)&=\frac{\Gamma(1-a-b)}{\Gamma(1-a)\Gamma(1-b)}\,{}_2F_1\left(a,b;c;x\right)
+\frac{\Gamma(a+b-1)}{\Gamma(a)\Gamma(b)}\,x^{1-c}\,{}_2F_1\left(1-b,1-a;2-a-b;x\right)~,
\end{align}
and hence it follows that $u^{\rm (I)}(x)$, presented in \ref{hypergcase2} becomes $u^{\rm (I)}(x)=\mathcal{A}(1-a-b)\,{}_2F_1\left(a,b;1;1-x\right)$. In the limit $x\to 0$ (corresponding to the asymptotic limit), we obtain
\begin{equation}\label{imp}
u^{\rm (I)}(x\to 0)=\mathcal{A}\left(\frac{\Gamma(2-a-b)}{\Gamma(1-a)\Gamma(1-b)}-\frac{\Gamma(a+b)}{\Gamma(a)\Gamma(b)} x^{1-c}\right)~,
\end{equation}
where, we have used the result that, $\lim_{x\to 0}{}_2F_1\left(a,b;c;x\right)=1$. Using the fact that $\Gamma(2-a-b)=(1-a-b)\Gamma(1-a-b)$, and $\Gamma(a+b)=(a+b-1)\Gamma(a+b-1)$, it follows that $u(x)$ in the above limit can be expressed as,  
\begin{align}
u^{\rm (I)}(x\to 0)&=\mathcal{A}(1-a-b)\frac{\Gamma(a+b-1)}{\Gamma(a)\Gamma(b)}x^{1-c}\left[1+\mathcal{F}^{\rm (I)}x^{c-1}\right]~,\label{response_gen}
\\
\mathcal{F}^{\rm (I)}&=\frac{\Gamma(1-a-b)\Gamma(a)\Gamma(b)}{\Gamma(1-a)\Gamma(1-b)\Gamma(a+b-1)}
=(a+b-1)\frac{\Gamma(a)^2 \Gamma(b)^2}{\Gamma(a + b)^2}\frac{\sin(\pi a)\sin(\pi b)}{\pi \sin[\pi(a + b)]}~.
\label{response_I}
\end{align}
In arriving at the above result, we have used the identity: $\Gamma(z)\Gamma(1-z)=\pi/\sin(\pi z)$. The above result for $\mathcal{F}^{\rm (I)}$ has been used extensively in the main text.

The second possibility for the constants $(a,b,c)$ is the following: $c \in \mathbb{Z}$ and $(a,b)$ are half-integers. In this case, the above expression for $\mathcal{F}$ does not work, since $\sin(\pi a) = \pm1 = \sin(\pi b)$, while $\sin[\pi(a + b)]=\sin(\pi c)=0$, leading to divergent $\mathcal{F}$. Thus, we have to use a different strategy. The complete solution of the hypergeometric differential equation in the present context, where, $c=n$, with $n\in \mathbb{Z}$ is [DLMF (15.10.8) \cite{NIST:DLMF}],
\begin{align}
u^{\rm (II)}(x)&=A\,{}_2F_1\left(a,b;n;x\right)+B\bigg\{{}_2F_1\left(a,b;n;x\right)\ln x
-\sum^{n-1}_{k=1}\frac{(n-1)!(k-1)!\Gamma(1-a)\Gamma(1-b)}{(n-k-1)!\Gamma(1-a+k)\Gamma(1-b+k)}(-x)^{-k} \nonumber
\\
&+\sum^{\infty}_{k=0} \frac{\Gamma(a+k)\Gamma(b+k)}{\Gamma(a)\Gamma(b)\, k!} x^k
\left[\psi(a+k)+\psi(b+k)-\psi(1+k)-\psi(n+k) \right] \bigg\}~.
\end{align}
Even though the first solution ${}_2F_1\left[a,b;n;x\right]$ has a $\log(1-x)$ piece, and hence is ill-behaved in the $x\to 1$ limit, the other solution is regular there. Since, the term ${}_2F_1\left(a,b;n;x\right)\ln x$ becomes $\log(1-x)\ln x \sim (1-x)\log(1-x)$, which is finite in the $x\to 1$ limit. Thus, regularity of $u(x)$ at $x=1$, demands $A = 0$. The resultant solution for $u(x)$, regular at $x=1$, in the $x\to 0$ limit becomes,
\begin{align}
u^{\rm (II)}(x\to 0)&=B\left[\ln(x)+(-1)^n x^{1-n}\frac{\Gamma(1-a)\Gamma(1-b)}{\Gamma(a)\Gamma(b)}(n-1)!(n-2)!+\cdots+\textrm{positive~powers~of~x}\right]
\nonumber
\\
&=B(-1)^n x^{1-n}\frac{\Gamma(1-a)\Gamma(1-b)}{\Gamma(a)\Gamma(b)}(n-1)!(n-2)!\left[1+\mathcal{F}^{\rm (II)}x^{n-1}\right]\,,\label{response_half}
\end{align}
where the quantity $\mathcal{F}^{\rm (II)}$ reads (using $n=c$),
\begin{equation}
\mathcal{F}^{\rm (II)}=(-1)^{c}\frac{\Gamma(a)^{2}\Gamma(b)^{2}}{(c-1)!(c-2)!}\frac{\sin(\pi a)\sin(\pi b)}{\pi^2} \ln x~.
\end{equation}
Finally, for all the constants in the hypergeometric differential equation satisfying $(a,b,c) \in \mathbb{Z}$, it follows that $\sin(\pi a)=0=\sin(\pi b)=\sin(\pi c)$. Therefore, \ref{response_gen} tells us that the corresponding quantity $\mathcal{F}^{\rm (III)}$ vanishes identically. This can also be seen from \ref{imp}, as the ratio $\{\Gamma(2-c)/\Gamma(1-a)\Gamma(1-b)\}$ vanishes, and hence the corresponding solution for the hypergeometric function in the $x\to 0$ limit becomes,
\begin{equation}\label{response_Int}
u^{\rm (III)}=-A\frac{\Gamma(a+b)}{\Gamma(a)\Gamma(b)}x^{1-c}\,,
\qquad 
\mathcal{F}^{\rm (III)}=0\,.
\end{equation}
This also shows that for $(a,b,c) \in \mathbb{Z}$, the corresponding quantity $\mathcal{F}^{\rm (III)}$ vanishes. These results will be used in the main text. 

\section{Solution of gravitational field equations in Lovelock gravity}\label{App:gravlove}

In this appendix, we derive the algebraic equation from which the solution of the gravitational field equations in Lovelock gravity can be derived. Even though a sketch of this calculation is provided in \cite{Takahashi:2010ye}, the details are missing there, which are not easy to grasp. Thus, for completeness and for the sake of the readers we are providing here the details. These results will also be used in the main text.  

The starting point is the Lovelock Lagrangian, which reads as,
\begin{equation}
L=\sum_{m=0}^{k}\frac{a_{m}}{2^{m}}\delta^{\lambda_{1} \sigma_{1}...\lambda_{m} \sigma_{m}}_{\rho_{1}\kappa_{1}...\rho_{m}\kappa_{m}}R^{\rho_{1}\kappa_{1}}_{\lambda_{1} \sigma_{1}}...R^{\rho_{m}\kappa_{m}}_{\lambda_{m} \sigma_{m}}~,
\end{equation}
where, $a_{m}$ is the coupling constant, $k$ should satisfy the following inequality: $d\geq 2k+1$. Note that the $m=0$ term describes the cosmological constant, with $a_{0}=2\Lambda$, $m=1$ term is general relativity with $a_{1}=1$, and $m=2$ is the Gauss-Bonnet term with $a_{2}=\alpha$, the Gauss-Bonnet coupling constant. Note that we have set $16\pi G=1$ in this appendix. Given the above Lagrangian, we can vary it with respect to the metric tensor and hence obtain the following field equations for Lovelock theories of gravity as,
\begin{eqnarray}\label{EoM_Lovelock}
-\sum_{m=1}^{k}\frac{a_{m}}{2^{m+1}}\delta^{\mu\lambda_{1} \sigma_{1}...\lambda_{m} \sigma_{m}}_{\nu\rho_{1}\kappa_{1}...\rho_{m}\kappa_{m}}R^{\rho_{1}\kappa_{1}}_{\lambda_{1} \sigma_{1}}...R^{\rho_{m}\kappa_{m}}_{\lambda_{m} \sigma_{m}}+\Lambda \delta^{\mu}_{\nu}=8 \pi T^{\mu}_{\nu}~.
\end{eqnarray}
The structure of the Equation of motion looks complicated, however, as we shall show, it gets simplified in the case of static and spherically symmetric BH spacetimes. For this case, we consider the following metric ansatz,
\begin{equation}
ds^{2}=-f(r)dt^2+\frac{dr^{2}}{f(r)}+r^2d\Omega^{2}_{d-2}~,
\end{equation}
for which only the following components of the Riemann tensor are non-zero,
\begin{eqnarray}\label{riemann_components}
R^{tr}_{tr}=-\frac{f''}{2}\,;
\quad
R^{tj}_{ti}=R^{rj}_{ri}=\frac{f'}{2r}\delta^{j}_{i}\,;
\quad
R^{mn}_{ij}=\bigg(\frac{(1-f)}{r^2}\bigg)\bigg(\delta^{m}_{i}\delta^{n}_{j}-\delta^{m}_{j}\delta^{n}_{i}\bigg)~.
\end{eqnarray}
Keeping future simplifications in mind, we define a new variable $\Psi$ as, $f=1-r^{2}\Psi(r)$ and shall solve for this variable. The field equations will have various components, but for the determination of $\Psi$ only the $(t,t)$ components of the field equation matter. Thus taking $\mu=t=\nu$ in \ref{EoM_Lovelock}, we obtain the corresponding field equation as,
\begin{align}
\sum_{m=1}^{k}\frac{4ma_{m}}{2^{m+1}}\delta^{trip_{2}q_{2}\cdots p_{m}q_{m}}_{trj m_{2}n_{2}\cdots m_{m}n_{m}}R^{m_{2}n_{2}}_{p_{2}q_{2}}\cdots R^{m_{m}n_{m}}_{p_{m} q_{m}}R^{rj}_{ri}
+\sum_{m=1}^{k}\frac{a_{m}}{2^{m+1}}\delta^{t i_{1}j_{1}\cdots i_{m}j_{m}}_{t m_{1}n_{1}\cdots m_{m}n_{m}}R^{m_{1}n_{1}}_{i_{1} j_{1}}\cdots R^{m_{m}n_{m}}_{i_{m} j_{m}}=8 \pi \rho+\Lambda~,
\end{align}
where, we have taken the energy momentum tensor in the field equation to be given by dust, with an energy density $\rho$ and vanishing pressure. Substituting all the non-trivial components of Riemann, as presented in \ref{riemann_components}, we obtain the following expression for the $(t,t)$ component of the field equations in terms of the quantity $\Psi$ and its derivatives, 
\begin{align}
-\sum_{m=1}^{k}\frac{4ma_{m}}{2^{m+1}}&\delta^{ip_{2}q_{2}\cdots p_{m}q_{m}}_{j m_{2}n_{2}\cdots m_{m}n_{m}}\delta^{j}_{i}\delta^{m_{2}n_{2}}_{p_{2}q_{2}}\cdots \delta^{m_{m}n_{m}}_{p_{m}q_{m}}\Psi^{m-1}\bigg(\frac{r\Psi'}{2}+\Psi\bigg)
\nonumber
\\
&-\sum_{m=1}^{k}\frac{a_{m}}{2^{m+1}}\delta^{i_{1}j_{1}\cdots i_{m}j_{m}}_{ m_{1}n_{1}\cdots m_{m}n_{m}}\delta^{m_{1}n_{1}}_{i_{1} j_{1}}\cdots \delta^{m_{m}n_{m}}_{i_{m} j_{m}}\Psi^{m}+\Lambda=-8 \pi\rho~.
\end{align}
Multiplying the above equation by $r^{d-2}$, and separating out the general relativistic term from the above Lovelock field equations, we obtain,
\begin{align}\label{reducedtt}
-r^{d-2}&\bigg[(d-2)\bigg(\frac{r \Psi'}{2}+\Psi\bigg)+\frac{(d-2)(d-3)}{2}\Psi\bigg]-\sum_{m=2}^{k} \frac{a_{m}}{2^{m+1}}\bigg[4m\delta^{i p_{2}q_{2}\cdots p_{m}q_{m}}_{j m_{2}n_{2}\cdots m_{m}n_{m}}\delta^{j}_{i}\delta^{m_{2}n_{2}}_{p_{2} q_{2}}\cdots \delta^{m_{m}n_{m}}_{p_{m} q_{m}}
\nonumber
\\
&\times\Psi^{m-1}\bigg(\frac{r\Psi'}{2}+\Psi\bigg)r^{d-2}+\delta^{i_{1}j_{1}\cdots i_{m}j_{m}}_{ p_{1}q_{1}\cdots p_{m}q_{m}}\delta^{p_{1}q_{1}}_{i_{1} j_{1}}\cdots \delta^{p_{m}q_{m}}_{i_{m} j_{m}}\Psi^{m}r^{d-2}\bigg]+\Lambda r^{d-2}=-8 \pi \rho r^{d-2}~.
\end{align}
Note that, the first term, which provides the contribution from general relativity, can be expressed as,
\begin{eqnarray}
~\bigg[(d-2)\bigg(\frac{r \Psi'}{2}+\Psi\bigg)+\frac{(d-2)(d-3)}{2}\Psi\bigg]r^{d-2}=\frac{(d-2)}{2}\frac{d}{dr}\bigg[r^{d-1}\Psi\bigg]~.
\end{eqnarray}
Moreover, the combination of Kronecker delta's can be decomposed as,
\begin{align}
\delta^{i_{1}j_{1}\cdots i_{m}j_{m}}_{m_{1}n_{1}\cdots m_{m}n_{m}}\delta^{m_{1}n_{1}}_{i_{1} j_{1}}=2\delta^{i_{1}j_{1}\cdots i_{m}j_{m}}_{m_{1}n_{1}\cdots m_{m}n_{m}}\delta^{m_{1}}_{i_{1}}\delta^{n_{1}}_{j_{1}}~,
\end{align}
and hence the field equation presented in \ref{reducedtt} becomes,
\begin{align}
-\frac{(d-2)}{2}\frac{d}{dr}\bigg[r^{d-1}\Psi\bigg]&-\sum_{m=2}\frac{a_{m}}{2^{m+1}}\bigg[4m2^{m-1}\Psi^{m-1}r^{d-2}\bigg(\frac{r\Psi'}{2}+\Psi\bigg)\Delta_{2m-1}
\nonumber
\\
&+2^{m}\Psi^{m}r^{d-2}\Delta_{2m}\bigg]+\Lambda r^{d-2}=-8 \pi \rho r^{d-2}~,
\end{align}
where, we have introduced the following definition:
\begin{eqnarray}
\Delta_{m}=\delta^{i_{1}i_{2}\cdots i_{m}}_{j_{1}j_{2}...j_{m}}\delta^{j_{1}}_{i_{1}}\delta^{j_{2}}_{i_{2}}\cdots\delta^{j_{m}}_{i_{m}}~.
\end{eqnarray}
The above quantity can be determined for each value of the Lovelock order $m$, and for the first two values of $m$, i.e., for Einstein-Hilbert and Gauss-Bonnet, we will obtain,
\begin{eqnarray}
\Delta_{1}=\delta^{i_{1}}_{j_{1}}\delta^{j_{1}}_{i_{1}}=(d-2)~;
\quad 
\Delta_{2}=\delta^{i_{1}i_{2}}_{j_{1}j_{2}}\delta^{j_{1}}_{i_{1}}\delta^{j_{2}}_{i_{2}}=(d-2)(d-3)~.
\end{eqnarray}
Thus, by induction, and for generic Lovelock order $m$, we obtain, $\Delta_{m}=(d-2)(d-3)\cdots(d-m-1)$. Therefore, the equation of motion for $\psi$ has been expressed as,
\begin{align}
-\frac{(d-2)}{2}\frac{d}{dr}\bigg[r^{d-1}\Psi\bigg]&-\sum_{m=2}^{k}\frac{a_{m}}{2^{m+1}}\bigg[2^{m+1}m(d-2)(d-3)\cdots(d-2m)\Psi^{m-1}\bigg(\frac{r\Psi'}{2}+\Psi\bigg)r^{d-2}
\nonumber
\\
&+2^{m}\Psi^{m}r^{d-2}(d-2)(d-3)\cdots(d-2m-1)\bigg]+\Lambda r^{d-2}=-8 \pi \rho r^{d-2}\,.
\end{align}
Simplifying further and then integrating the above equation over the radial coordinate $r$, we obtain,
\begin{align}
r^{d-1}\Psi+\sum_{m=2}^{k}\frac{a_{m}(d-3)!}{(d-2m-1)!}r^{d-1}\Psi^{m}-\frac{2 \Lambda}{(d-2)}\frac{r^{d-1}}{(d-1)}=\frac{16 \pi}{(d-2)}\int \rho r^{d-2}dr\,.
\end{align}
Taking a cue from four-dimensional general relativity, we define the mass of the above geometry as the following integral of the density function $\rho(r)$,
\begin{equation}
M=\frac{2 \pi^{(d-1)/2}}{\Gamma\bigg(\frac{d-1}{2}\bigg)}\int \rho r^{d-2} dr=S_{d-2}\int \rho r^{d-2} dr\,,
\end{equation}
where, $S_{d-2}$ is the area of the $(d-2)$ dimensional unit sphere, given by $S_{d-2}=\{2\pi^{(d-1)/2}/\Gamma(d-1/2)\}$, which for $d=4$, reduces to $S_{2}=4\pi$, as expected. Therefore, we finally obtain the desired algebraic equation for the determination of the metric function $\Psi(r)$, 
\begin{eqnarray}\label{PsiEq}
\frac{2\mathbb{M}}{r^{d-1}}&=&\Psi+\sum_{m=2}\frac{a_{m}(d-3)!}{(d-2m-1)!}\Psi^{m}-\frac{2 \Lambda}{(d-2)(d-1)}\nonumber\\&=&\sum_{m=2}\Bigg[a_{m}\bigg\{\prod_{p=1}^{2m-2}(d-2-p)\bigg\}\Psi^{m}\bigg]+\Psi-\frac{2 \Lambda}{(d-2)(d-1)}~,
\end{eqnarray}
where, we have to use the following identity,
\begin{equation}
\frac{(d-3)!}{(d-2m-1)!}=(d-3)...(d-2m)=\prod_{p=1}^{2m-2}(d-2-p) ~.
\end{equation}
along with, we have introduced the rescaled mass $\mathbb{M}$, which is related to the mass M of the density distribution, as,
\begin{equation}
M=\frac{2 \pi^{\frac{(d-1)}{2}}}{\Gamma\bigg(\frac{d-1}{2}\bigg)}\frac{(d-2)\mathbb{M}}{8\pi}=\frac{S_{d-2}(d-2)\mathbb{M}}{8\pi}~.
\end{equation}
considering $\Lambda=0$, i.e., asymptotically flat geometries, we have $\Psi=(2\mathbb{M}/r^{d-1})$, and hence, $f(r)=1-(2\mathbb{M}/r^{d-3})$, for general relativity. For pure Lovelock gravity of order $N$, we obtain, $\Psi^{N}=(2\mathbb{M}/r^{d-1})$, such that $f=1-(2\mathbb{M}/r^{d-2N-1})^{1/N}$. This is one of the results that we have used in the main text. Here, we have taken the coupling constant $a_{N}$, such that $a_{N}(d-3)\cdots (d-2N)=1$. Finally, for Einstein-Gauss-Bonnet gravity, we have $a_{2}=\alpha$, and hence from \ref{PsiEq} it follows that the $\Psi$ satisfies the following equation,
\begin{equation}
\frac{2\mathbb{M}}{r^{d-1}}=\Psi+\alpha(d-3)(d-4)\Psi^{2}=\Psi+\tilde{\alpha}\Psi^{2}~,
\end{equation}
where, $\tilde{\alpha}$ corresponds to the rescaled Gauss-Bonnet parameter. The above quadratic equation can be immediately solved, with the boundary condition, that $\Psi$ should vanish asymptotically, yielding, 
\begin{equation}
\Psi=-\frac{1}{2\tilde{\alpha}}+\frac{1}{2\tilde{\alpha}}\sqrt{1-\frac{8\tilde{\alpha}\mathbb{M}}{r^{d-1}}}\,,
\end{equation}
which is equal to, 
\begin{equation}
\Psi=-\frac{1}{2\tilde{\alpha}}\pm \frac{1}{2\tilde{\alpha}}\sqrt{1-\frac{64 \pi \tilde{\alpha}M}{S_{d-2}(d-2)r^{d-1}}}\,. 
\end{equation}
This result has been used in the main text. 

\bibliography{reference.bib}

\providecommand{\href}[2]{#2}\begingroup\raggedright\begin{thebibliography}{10}

\bibitem{Kerr:1963ud}
R.~P. Kerr, ``{Gravitational field of a spinning mass as an example of
  algebraically special metrics},''
  \href{http://dx.doi.org/10.1103/PhysRevLett.11.237}{{\em Phys. Rev. Lett.}
  {\bfseries 11} (1963) 237--238}.

\bibitem{Newman:1965my}
E.~T. Newman, R.~Couch, K.~Chinnapared, A.~Exton, A.~Prakash, and R.~Torrence,
  ``{Metric of a Rotating, Charged Mass},''
  \href{http://dx.doi.org/10.1063/1.1704351}{{\em J. Math. Phys.} {\bfseries 6}
  (1965) 918--919}.

\bibitem{Israel:1967wq}
W.~Israel, ``{Event horizons in static vacuum space-times},''
  \href{http://dx.doi.org/10.1103/PhysRev.164.1776}{{\em Phys. Rev.} {\bfseries
  164} (1967) 1776--1779}.

\bibitem{Carter:1971zc}
B.~Carter, ``{Axisymmetric Black Hole Has Only Two Degrees of Freedom},''
  \href{http://dx.doi.org/10.1103/PhysRevLett.26.331}{{\em Phys. Rev. Lett.}
  {\bfseries 26} (1971) 331--333}.

\bibitem{Bekenstein:1971hc}
J.~D. Bekenstein, ``{Nonexistence of baryon number for static black holes},''
  \href{http://dx.doi.org/10.1103/PhysRevD.5.1239}{{\em Phys. Rev. D}
  {\bfseries 5} (1972) 1239--1246}.

\bibitem{Bekenstein:1995un}
J.~D. Bekenstein, ``{Novel
  \textquoteleft{}\textquoteleft{}no-scalar-hair\textquoteright{}\textquoteright{}
  theorem for black holes},''
  \href{http://dx.doi.org/10.1103/PhysRevD.51.R6608}{{\em Phys. Rev. D}
  {\bfseries 51} no.~12, (1995) R6608}.

\bibitem{Hui:2012qt}
L.~Hui and A.~Nicolis, ``{No-Hair Theorem for the Galileon},''
  \href{http://dx.doi.org/10.1103/PhysRevLett.110.241104}{{\em Phys. Rev.
  Lett.} {\bfseries 110} (2013) 241104},
  \href{http://arxiv.org/abs/1202.1296}{{\ttfamily arXiv:1202.1296 [hep-th]}}.

\bibitem{Jacobson:1999vr}
T.~Jacobson, ``{Primordial black hole evolution in tensor scalar cosmology},''
  \href{http://dx.doi.org/10.1103/PhysRevLett.83.2699}{{\em Phys. Rev. Lett.}
  {\bfseries 83} (1999) 2699--2702},
  \href{http://arxiv.org/abs/astro-ph/9905303}{{\ttfamily
  arXiv:astro-ph/9905303}}.

\bibitem{Horbatsch:2011ye}
M.~W. Horbatsch and C.~P. Burgess, ``{Cosmic Black-Hole Hair Growth and Quasar
  OJ287},'' \href{http://dx.doi.org/10.1088/1475-7516/2012/05/010}{{\em JCAP}
  {\bfseries 05} (2012) 010}, \href{http://arxiv.org/abs/1111.4009}{{\ttfamily
  arXiv:1111.4009 [gr-qc]}}.

\bibitem{Hui:2019aqm}
L.~Hui, D.~Kabat, X.~Li, L.~Santoni, and S.~S.~C. Wong, ``{Black Hole Hair from
  Scalar Dark Matter},''
  \href{http://dx.doi.org/10.1088/1475-7516/2019/06/038}{{\em JCAP} {\bfseries
  06} (2019) 038}, \href{http://arxiv.org/abs/1904.12803}{{\ttfamily
  arXiv:1904.12803 [gr-qc]}}.

\bibitem{Clough:2019jpm}
K.~Clough, P.~G. Ferreira, and M.~Lagos, ``{Growth of massive scalar hair
  around a Schwarzschild black hole},''
  \href{http://dx.doi.org/10.1103/PhysRevD.100.063014}{{\em Phys. Rev. D}
  {\bfseries 100} no.~6, (2019) 063014},
  \href{http://arxiv.org/abs/1904.12783}{{\ttfamily arXiv:1904.12783 [gr-qc]}}.

\bibitem{Penrose:1969pc}
R.~Penrose, ``{Gravitational collapse: The role of general relativity},''
  \href{http://dx.doi.org/10.1023/A:1016578408204}{{\em Riv. Nuovo Cim.}
  {\bfseries 1} (1969) 252--276}.

\bibitem{1971JETPL..14..180Z}
Y.~B. {Zel'Dovich}, ``{Generation of Waves by a Rotating Body},'' {\em Soviet
  Journal of Experimental and Theoretical Physics Letters} {\bfseries 14}
  (Aug., 1971) 180.

\bibitem{1972JETP...35.1085Z}
Y.~B. {Zel'Dovich}, ``{Amplification of Cylindrical Electromagnetic Waves
  Reflected from a Rotating Body},'' {\em Soviet Journal of Experimental and
  Theoretical Physics} {\bfseries 35} (Jan., 1972) 1085.

\bibitem{1972BAPS...17..472M}
C.~{Misner}, ``{Stability of Kerr black holes against scalar perturbations},''
  in {\em Bulletin of the American Physical Society}, vol.~17, p.~472.
\newblock Dec., 1972.

\bibitem{Bardeen:1972fi}
J.~M. Bardeen, W.~H. Press, and S.~A. Teukolsky, ``{Rotating black holes:
  Locally nonrotating frames, energy extraction, and scalar synchrotron
  radiation},'' \href{http://dx.doi.org/10.1086/151796}{{\em Astrophys. J.}
  {\bfseries 178} (1972) 347}.

\bibitem{Press:1972zz}
W.~H. Press and S.~A. Teukolsky, ``{Floating Orbits, Superradiant Scattering
  and the Black-hole Bomb},'' \href{http://dx.doi.org/10.1038/238211a0}{{\em
  Nature} {\bfseries 238} (1972) 211--212}.

\bibitem{Starobinsky:1973aij}
A.~A. Starobinsky, ``{Amplification of waves reflected from a rotating ''black
  hole''.},'' {\em Sov. Phys. JETP} {\bfseries 37} no.~1, (1973) 28--32.

\bibitem{Teukolsky:1974yv}
S.~A. Teukolsky and W.~H. Press, ``{Perturbations of a rotating black hole. III
  - Interaction of the hole with gravitational and electromagnet ic
  radiation},'' \href{http://dx.doi.org/10.1086/153180}{{\em Astrophys. J.}
  {\bfseries 193} (1974) 443--461}.

\bibitem{Arvanitaki:2010sy}
A.~Arvanitaki and S.~Dubovsky, ``{Exploring the String Axiverse with Precision
  Black Hole Physics},''
  \href{http://dx.doi.org/10.1103/PhysRevD.83.044026}{{\em Phys. Rev. D}
  {\bfseries 83} (2011) 044026},
  \href{http://arxiv.org/abs/1004.3558}{{\ttfamily arXiv:1004.3558 [hep-th]}}.

\bibitem{Endlich:2016jgc}
S.~Endlich and R.~Penco, ``{A Modern Approach to Superradiance},''
  \href{http://dx.doi.org/10.1007/JHEP05(2017)052}{{\em JHEP} {\bfseries 05}
  (2017) 052}, \href{http://arxiv.org/abs/1609.06723}{{\ttfamily
  arXiv:1609.06723 [hep-th]}}.

\bibitem{Baumann:2019eav}
D.~Baumann, H.~S. Chia, J.~Stout, and L.~ter Haar, ``{The Spectra of
  Gravitational Atoms},''
  \href{http://dx.doi.org/10.1088/1475-7516/2019/12/006}{{\em JCAP} {\bfseries
  12} (2019) 006}, \href{http://arxiv.org/abs/1908.10370}{{\ttfamily
  arXiv:1908.10370 [gr-qc]}}.

\bibitem{Cardoso:2017cfl}
V.~Cardoso, E.~Franzin, A.~Maselli, P.~Pani, and G.~Raposo, ``{Testing
  strong-field gravity with tidal Love numbers},''
  \href{http://dx.doi.org/10.1103/PhysRevD.95.084014}{{\em Phys. Rev. D}
  {\bfseries 95} no.~8, (2017) 084014},
  \href{http://arxiv.org/abs/1701.01116}{{\ttfamily arXiv:1701.01116 [gr-qc]}}.
  [Addendum: Phys.Rev.D 95, 089901 (2017)].

\bibitem{Chakraborty:2025wvs}
S.~Chakraborty, V.~De~Luca, L.~Gualtieri, and P.~Pani, ``{Dynamical Love
  numbers of black holes: theory and gravitational waveforms},''
  \href{http://arxiv.org/abs/2507.22994}{{\ttfamily arXiv:2507.22994 [gr-qc]}}.

\bibitem{Hinderer:2007mb}
T.~Hinderer, ``{Tidal Love numbers of neutron stars},''
  \href{http://dx.doi.org/10.1086/533487}{{\em Astrophys. J.} {\bfseries 677}
  (2008) 1216--1220}, \href{http://arxiv.org/abs/0711.2420}{{\ttfamily
  arXiv:0711.2420 [astro-ph]}}. [Erratum: Astrophys.J. 697, 964 (2009)].

\bibitem{hinderer2009tidal-327}
T.~Hinderer, B.~D. Lackey, R.~N. Lang, and J.~S. Read, ``Tidal deformability of
  neutron stars with realistic equations of state and their gravitational wave
  signatures in binary inspiral,''
  \href{http://dx.doi.org/10.48550/arxiv.0911.3535}{{\em {arXiv}} (2009) },
  \href{http://arxiv.org/abs/0911.3535}{{\ttfamily 0911.3535}}.

\bibitem{yagi2013multipole-5d7}
K.~Yagi, ``Multipole love relations,''
  \href{http://dx.doi.org/10.48550/arxiv.1311.0872}{{\em {arXiv}} (2013) },
  \href{http://arxiv.org/abs/1311.0872}{{\ttfamily 1311.0872}}.

\bibitem{Yagi:2016bkt}
K.~Yagi and N.~Yunes, ``{Approximate Universal Relations for Neutron Stars and
  Quark Stars},'' \href{http://dx.doi.org/10.1016/j.physrep.2017.03.002}{{\em
  Phys. Rept.} {\bfseries 681} (2017) 1--72},
  \href{http://arxiv.org/abs/1608.02582}{{\ttfamily arXiv:1608.02582 [gr-qc]}}.

\bibitem{LIGOScientific:2017vwq}
{\bfseries LIGO Scientific, Virgo} Collaboration, B.~P. Abbott {\em et~al.},
  ``{GW170817: Observation of Gravitational Waves from a Binary Neutron Star
  Inspiral},'' \href{http://dx.doi.org/10.1103/PhysRevLett.119.161101}{{\em
  Phys. Rev. Lett.} {\bfseries 119} no.~16, (2017) 161101},
  \href{http://arxiv.org/abs/1710.05832}{{\ttfamily arXiv:1710.05832 [gr-qc]}}.

\bibitem{Binnington:2009bb}
T.~Binnington and E.~Poisson, ``{Relativistic theory of tidal Love numbers},''
  \href{http://dx.doi.org/10.1103/PhysRevD.80.084018}{{\em Phys. Rev. D}
  {\bfseries 80} (2009) 084018},
  \href{http://arxiv.org/abs/0906.1366}{{\ttfamily arXiv:0906.1366 [gr-qc]}}.

\bibitem{Damour:2009vw}
T.~Damour and A.~Nagar, ``{Relativistic tidal properties of neutron stars},''
  \href{http://dx.doi.org/10.1103/PhysRevD.80.084035}{{\em Phys. Rev. D}
  {\bfseries 80} (2009) 084035},
  \href{http://arxiv.org/abs/0906.0096}{{\ttfamily arXiv:0906.0096 [gr-qc]}}.

\bibitem{Takahashi:2010ye}
T.~Takahashi and J.~Soda, ``{Master Equations for Gravitational Perturbations
  of Static Lovelock Black Holes in Higher Dimensions},''
  \href{http://dx.doi.org/10.1143/PTP.124.911}{{\em Prog. Theor. Phys.}
  {\bfseries 124} (2010) 911--924},
  \href{http://arxiv.org/abs/1008.1385}{{\ttfamily arXiv:1008.1385 [gr-qc]}}.

\bibitem{Ishibashi:2003ap}
A.~Ishibashi and H.~Kodama, ``{Stability of higher dimensional Schwarzschild
  black holes},'' \href{http://dx.doi.org/10.1143/PTP.110.901}{{\em Prog.
  Theor. Phys.} {\bfseries 110} (2003) 901--919},
  \href{http://arxiv.org/abs/hep-th/0305185}{{\ttfamily arXiv:hep-th/0305185}}.

\bibitem{Hui:2020xxx}
L.~Hui, A.~Joyce, R.~Penco, L.~Santoni, and A.~R. Solomon, ``{Static response
  and Love numbers of Schwarzschild black holes},''
  \href{http://dx.doi.org/10.1088/1475-7516/2021/04/052}{{\em JCAP} {\bfseries
  04} (2021) 052}, \href{http://arxiv.org/abs/2010.00593}{{\ttfamily
  arXiv:2010.00593 [hep-th]}}.

\bibitem{charalambous2023scalar-ba1}
P.~Charalambous and M.~M. Ivanov, ``Scalar love numbers and love symmetries of
  5-dimensional myers-perry black holes,''
  \href{http://dx.doi.org/10.48550/arxiv.2303.16036}{{\em {arXiv}} (2023) },
  \href{http://arxiv.org/abs/2303.16036}{{\ttfamily 2303.16036}}.

\bibitem{Fang:2005qq}
H.~Fang and G.~Lovelace, ``{Tidal coupling of a Schwarzschild black hole and
  circularly orbiting moon},''
  \href{http://dx.doi.org/10.1103/PhysRevD.72.124016}{{\em Phys. Rev. D}
  {\bfseries 72} (2005) 124016},
  \href{http://arxiv.org/abs/gr-qc/0505156}{{\ttfamily arXiv:gr-qc/0505156}}.

\bibitem{Kol:2011vg}
B.~Kol and M.~Smolkin, ``{Black hole stereotyping: Induced gravito-static
  polarization},'' \href{http://dx.doi.org/10.1007/JHEP02(2012)010}{{\em JHEP}
  {\bfseries 02} (2012) 010}, \href{http://arxiv.org/abs/1110.3764}{{\ttfamily
  arXiv:1110.3764 [hep-th]}}.

\bibitem{Chakrabarti:2013lua}
S.~Chakrabarti, T.~Delsate, and J.~Steinhoff, ``{New perspectives on neutron
  star and black hole spectroscopy and dynamic tides},''
  \href{http://arxiv.org/abs/1304.2228}{{\ttfamily arXiv:1304.2228 [gr-qc]}}.

\bibitem{Gurlebeck:2015xpa}
N.~G\"urlebeck, ``{No-hair theorem for Black Holes in Astrophysical
  Environments},'' \href{http://dx.doi.org/10.1103/PhysRevLett.114.151102}{{\em
  Phys. Rev. Lett.} {\bfseries 114} no.~15, (2015) 151102},
  \href{http://arxiv.org/abs/1503.03240}{{\ttfamily arXiv:1503.03240 [gr-qc]}}.

\bibitem{hui2022ladder-678}
L.~Hui, A.~Joyce, R.~Penco, L.~Santoni, and A.~R. Solomon, ``Ladder symmetries
  of black holes. implications for love numbers and no-hair theorems,''
  \href{http://dx.doi.org/10.1088/1475-7516/2022/01/032}{{\em Journal of
  Cosmology and Astroparticle Physics} {\bfseries 2022} no.~01, (2022) 032},
  \href{http://arxiv.org/abs/2105.01069}{{\ttfamily 2105.01069}}.

\bibitem{achour2022hidden-8c7}
J.~B. Achour, E.~R. Livine, S.~Mukohyama, and J.-P. Uzan, ``Hidden symmetry of
  the static response of black holes: Applications to love numbers,''
  \href{http://dx.doi.org/10.48550/arxiv.2202.12828}{{\em {arXiv}} (2022) },
  \href{http://arxiv.org/abs/2202.12828}{{\ttfamily 2202.12828}}.

\bibitem{charalambous2021hidden-5e0}
P.~Charalambous, S.~Dubovsky, and M.~M. Ivanov, ``Hidden symmetry of vanishing
  love numbers,'' \href{http://dx.doi.org/10.1103/physrevlett.127.101101}{{\em
  Physical Review Letters} {\bfseries 127} no.~10, (2021) 101101},
  \href{http://arxiv.org/abs/2103.01234}{{\ttfamily 2103.01234}}.

\bibitem{ivanov2023vanishing-9aa}
M.~M. Ivanov and Z.~Zhou, ``Vanishing of black hole tidal love numbers from
  scattering amplitudes,''
  \href{http://dx.doi.org/10.1103/physrevlett.130.091403}{{\em Physical Review
  Letters} {\bfseries 130} no.~9, (2023) 091403},
  \href{http://arxiv.org/abs/2209.14324}{{\ttfamily 2209.14324}}.

\bibitem{creci2021tidal-42e}
G.~Creci, T.~Hinderer, and J.~Steinhoff, ``Tidal response from scattering and
  the role of analytic continuation,''
  \href{http://dx.doi.org/10.48550/arxiv.2108.03385}{{\em {arXiv}} (2021) },
  \href{http://arxiv.org/abs/2108.03385}{{\ttfamily 2108.03385}}.

\bibitem{Charalambous:2021mea}
P.~Charalambous, S.~Dubovsky, and M.~M. Ivanov, ``{On the Vanishing of Love
  Numbers for Kerr Black Holes},''
  \href{http://dx.doi.org/10.1007/JHEP05(2021)038}{{\em JHEP} {\bfseries 05}
  (2021) 038}, \href{http://arxiv.org/abs/2102.08917}{{\ttfamily
  arXiv:2102.08917 [hep-th]}}.

\bibitem{Cardoso:2019vof}
V.~Cardoso, L.~Gualtieri, and C.~J. Moore, ``{Gravitational waves and higher
  dimensions: Love numbers and Kaluza-Klein excitations},''
  \href{http://dx.doi.org/10.1103/PhysRevD.100.124037}{{\em Phys. Rev. D}
  {\bfseries 100} no.~12, (2019) 124037},
  \href{http://arxiv.org/abs/1910.09557}{{\ttfamily arXiv:1910.09557 [gr-qc]}}.

\bibitem{Emparan:2017qxd}
R.~Emparan, A.~Fernandez-Pique, and R.~Luna, ``{Geometric polarization of
  plasmas and Love numbers of AdS black branes},''
  \href{http://dx.doi.org/10.1007/JHEP09(2017)150}{{\em JHEP} {\bfseries 09}
  (2017) 150}, \href{http://arxiv.org/abs/1707.02777}{{\ttfamily
  arXiv:1707.02777 [hep-th]}}.

\bibitem{Cardoso:2018ptl}
V.~Cardoso, M.~Kimura, A.~Maselli, and L.~Senatore, ``{Black Holes in an
  Effective Field Theory Extension of General Relativity},''
  \href{http://dx.doi.org/10.1103/PhysRevLett.121.251105}{{\em Phys. Rev.
  Lett.} {\bfseries 121} no.~25, (2018) 251105},
  \href{http://arxiv.org/abs/1808.08962}{{\ttfamily arXiv:1808.08962 [gr-qc]}}.
  [Erratum: Phys.Rev.Lett. 131, 109903 (2023)].

\bibitem{DeLuca:2022tkm}
V.~De~Luca, J.~Khoury, and S.~S.~C. Wong, ``{Implications of the weak gravity
  conjecture for tidal Love numbers of black holes},''
  \href{http://dx.doi.org/10.1103/PhysRevD.108.044066}{{\em Phys. Rev. D}
  {\bfseries 108} no.~4, (2023) 044066},
  \href{http://arxiv.org/abs/2211.14325}{{\ttfamily arXiv:2211.14325
  [hep-th]}}.

\bibitem{Nair:2024mya}
S.~Nair, S.~Chakraborty, and S.~Sarkar, ``{Asymptotically de Sitter black holes
  have nonzero tidal Love numbers},''
  \href{http://dx.doi.org/10.1103/PhysRevD.109.064025}{{\em Phys. Rev. D}
  {\bfseries 109} no.~6, (2024) 064025},
  \href{http://arxiv.org/abs/2401.06467}{{\ttfamily arXiv:2401.06467 [gr-qc]}}.

\bibitem{Nair:2022xfm}
S.~Nair, S.~Chakraborty, and S.~Sarkar, ``{Dynamical Love numbers for area
  quantized black holes},''
  \href{http://dx.doi.org/10.1103/PhysRevD.107.124041}{{\em Phys. Rev. D}
  {\bfseries 107} no.~12, (2023) 124041},
  \href{http://arxiv.org/abs/2208.06235}{{\ttfamily arXiv:2208.06235 [gr-qc]}}.

\bibitem{DeLuca:2024ufn}
V.~De~Luca, A.~Garoffolo, J.~Khoury, and M.~Trodden, ``{Tidal Love numbers and
  Green\textquoteright{}s functions in black hole spacetimes},''
  \href{http://dx.doi.org/10.1103/PhysRevD.110.064081}{{\em Phys. Rev. D}
  {\bfseries 110} no.~6, (2024) 064081},
  \href{http://arxiv.org/abs/2407.07156}{{\ttfamily arXiv:2407.07156 [gr-qc]}}.

\bibitem{Bhatt:2024mvr}
R.~P. Bhatt and C.~Singha, ``{Scalar tidal response of a rotating BTZ black
  hole},'' \href{http://dx.doi.org/10.1007/JHEP11(2024)154}{{\em JHEP}
  {\bfseries 11} (2024) 154}, \href{http://arxiv.org/abs/2407.09470}{{\ttfamily
  arXiv:2407.09470 [gr-qc]}}.

\bibitem{Bhatt:2023zsy}
R.~P. Bhatt, S.~Chakraborty, and S.~Bose, ``{Addressing issues in defining the
  Love numbers for black holes},''
  \href{http://dx.doi.org/10.1103/PhysRevD.108.084013}{{\em Phys. Rev. D}
  {\bfseries 108} no.~8, (2023) 084013},
  \href{http://arxiv.org/abs/2306.13627}{{\ttfamily arXiv:2306.13627 [gr-qc]}}.

\bibitem{Bhatt:2024yyz}
R.~P. Bhatt, S.~Chakraborty, and S.~Bose, ``{Rotating black holes experience
  dynamical tides},''
  \href{http://dx.doi.org/10.1103/PhysRevD.111.L041504}{{\em Phys. Rev. D}
  {\bfseries 111} no.~4, (2025) L041504},
  \href{http://arxiv.org/abs/2406.09543}{{\ttfamily arXiv:2406.09543 [gr-qc]}}.

\bibitem{katagiri2024relativistic-5c0}
T.~Katagiri, K.~Yagi, and V.~Cardoso, ``On relativistic dynamical tides:
  subtleties and calibration,''
  \href{http://dx.doi.org/10.48550/arxiv.2409.18034}{{\em {arXiv}} (2024) },
  \href{http://arxiv.org/abs/2409.18034}{{\ttfamily 2409.18034}}.

\bibitem{saketh2023dynamical-b30}
M.~V.~S. Saketh, Z.~Zhou, and M.~M. Ivanov, ``Dynamical tidal response of kerr
  black holes from scattering amplitudes,''
  \href{http://dx.doi.org/10.48550/arxiv.2307.10391}{{\em {arXiv}} (2023) },
  \href{http://arxiv.org/abs/2307.10391}{{\ttfamily 2307.10391}}.

\bibitem{Chakraborty:2023zed}
S.~Chakraborty, E.~Maggio, M.~Silvestrini, and P.~Pani, ``{Dynamical tidal Love
  numbers of Kerr-like compact objects},''
  \href{http://dx.doi.org/10.1103/PhysRevD.110.084042}{{\em Phys. Rev. D}
  {\bfseries 110} no.~8, (2024) 084042},
  \href{http://arxiv.org/abs/2310.06023}{{\ttfamily arXiv:2310.06023 [gr-qc]}}.

\bibitem{Kehagias:2024rtz}
A.~Kehagias and A.~Riotto, ``{Black holes in a gravitational field: the
  non-linear static love number of Schwarzschild black holes vanishes},''
  \href{http://dx.doi.org/10.1088/1475-7516/2025/05/039}{{\em JCAP} {\bfseries
  05} (2025) 039}, \href{http://arxiv.org/abs/2410.11014}{{\ttfamily
  arXiv:2410.11014 [gr-qc]}}.

\bibitem{LeTiec:2020bos}
A.~Le~Tiec, M.~Casals, and E.~Franzin, ``{Tidal Love Numbers of Kerr Black
  Holes},'' \href{http://dx.doi.org/10.1103/PhysRevD.103.084021}{{\em Phys.
  Rev. D} {\bfseries 103} no.~8, (2021) 084021},
  \href{http://arxiv.org/abs/2010.15795}{{\ttfamily arXiv:2010.15795 [gr-qc]}}.

\bibitem{LeTiec:2020spy}
A.~Le~Tiec and M.~Casals, ``{Spinning Black Holes Fall in Love},''
  \href{http://dx.doi.org/10.1103/PhysRevLett.126.131102}{{\em Phys. Rev.
  Lett.} {\bfseries 126} no.~13, (2021) 131102},
  \href{http://arxiv.org/abs/2007.00214}{{\ttfamily arXiv:2007.00214 [gr-qc]}}.

\bibitem{Gounis:2024hcm}
L.~R. Gounis, A.~Kehagias, and A.~Riotto, ``{The vanishing of the non-linear
  static love number of Kerr black holes and the role of symmetries},''
  \href{http://dx.doi.org/10.1088/1475-7516/2025/03/002}{{\em JCAP} {\bfseries
  03} (2025) 002}, \href{http://arxiv.org/abs/2412.08249}{{\ttfamily
  arXiv:2412.08249 [gr-qc]}}.

\bibitem{Pani:2015hfa}
P.~Pani, L.~Gualtieri, A.~Maselli, and V.~Ferrari, ``{Tidal deformations of a
  spinning compact object},''
  \href{http://dx.doi.org/10.1103/PhysRevD.92.024010}{{\em Phys. Rev. D}
  {\bfseries 92} no.~2, (2015) 024010},
  \href{http://arxiv.org/abs/1503.07365}{{\ttfamily arXiv:1503.07365 [gr-qc]}}.

\bibitem{Pani:2015nua}
P.~Pani, L.~Gualtieri, and V.~Ferrari, ``{Tidal Love numbers of a slowly
  spinning neutron star},''
  \href{http://dx.doi.org/10.1103/PhysRevD.92.124003}{{\em Phys. Rev. D}
  {\bfseries 92} no.~12, (2015) 124003},
  \href{http://arxiv.org/abs/1509.02171}{{\ttfamily arXiv:1509.02171 [gr-qc]}}.

\bibitem{Landry:2015zfa}
P.~Landry and E.~Poisson, ``{Tidal deformation of a slowly rotating material
  body. External metric},''
  \href{http://dx.doi.org/10.1103/PhysRevD.91.104018}{{\em Phys. Rev. D}
  {\bfseries 91} (2015) 104018},
  \href{http://arxiv.org/abs/1503.07366}{{\ttfamily arXiv:1503.07366 [gr-qc]}}.

\bibitem{Landry:2015cva}
P.~Landry and E.~Poisson, ``{Gravitomagnetic response of an irrotational body
  to an applied tidal field},''
  \href{http://dx.doi.org/10.1103/PhysRevD.91.104026}{{\em Phys. Rev. D}
  {\bfseries 91} no.~10, (2015) 104026},
  \href{http://arxiv.org/abs/1504.06606}{{\ttfamily arXiv:1504.06606 [gr-qc]}}.

\bibitem{Landry:2017piv}
P.~Landry, ``{Tidal deformation of a slowly rotating material body: Interior
  metric and Love numbers},''
  \href{http://dx.doi.org/10.1103/PhysRevD.95.124058}{{\em Phys. Rev. D}
  {\bfseries 95} no.~12, (2017) 124058},
  \href{http://arxiv.org/abs/1703.08168}{{\ttfamily arXiv:1703.08168 [gr-qc]}}.

\bibitem{Poisson:2020mdi}
E.~Poisson, ``{Gravitomagnetic Love tensor of a slowly rotating body:
  post-Newtonian theory},''
  \href{http://dx.doi.org/10.1103/PhysRevD.102.064059}{{\em Phys. Rev. D}
  {\bfseries 102} no.~6, (2020) 064059},
  \href{http://arxiv.org/abs/2007.01678}{{\ttfamily arXiv:2007.01678 [gr-qc]}}.

\bibitem{Chia:2020yla}
H.~S. Chia, ``{Tidal deformation and dissipation of rotating black holes},''
  \href{http://dx.doi.org/10.1103/PhysRevD.104.024013}{{\em Phys. Rev. D}
  {\bfseries 104} no.~2, (2021) 024013},
  \href{http://arxiv.org/abs/2010.07300}{{\ttfamily arXiv:2010.07300 [gr-qc]}}.

\bibitem{Bhatt:2024rpx}
R.~P. Bhatt, S.~Chakraborty, and S.~Bose, ``{Response of a Kerr black hole to a
  generic tidal perturbation},''
  \href{http://arxiv.org/abs/2412.15117}{{\ttfamily arXiv:2412.15117 [gr-qc]}}.

\bibitem{Goldberger:2020fot}
W.~D. Goldberger, J.~Li, and I.~Z. Rothstein, ``{Non-conservative effects on
  spinning black holes from world-line effective field theory},''
  \href{http://dx.doi.org/10.1007/JHEP06(2021)053}{{\em JHEP} {\bfseries 06}
  (2021) 053}, \href{http://arxiv.org/abs/2012.14869}{{\ttfamily
  arXiv:2012.14869 [hep-th]}}.

\bibitem{Kobayashi:2019hrl}
T.~Kobayashi, ``{Horndeski theory and beyond: a review},''
  \href{http://dx.doi.org/10.1088/1361-6633/ab2429}{{\em Rept. Prog. Phys.}
  {\bfseries 82} no.~8, (2019) 086901},
  \href{http://arxiv.org/abs/1901.07183}{{\ttfamily arXiv:1901.07183 [gr-qc]}}.

\bibitem{Padmanabhan:2013xyr}
T.~Padmanabhan and D.~Kothawala, ``{Lanczos-Lovelock models of gravity},''
  \href{http://dx.doi.org/10.1016/j.physrep.2013.05.007}{{\em Phys. Rept.}
  {\bfseries 531} (2013) 115--171},
  \href{http://arxiv.org/abs/1302.2151}{{\ttfamily arXiv:1302.2151 [gr-qc]}}.

\bibitem{Diedrichs:2025vhv}
R.~F. Diedrichs, S.~Tsujikawa, and K.~Yagi, ``{Tidal Love numbers of neutron
  stars in Horndeski theories},''
  \href{http://dx.doi.org/10.1103/cmb4-chn3}{{\em Phys. Rev. D} {\bfseries 112}
  no.~4, (2025) 044023}, \href{http://arxiv.org/abs/2501.07998}{{\ttfamily
  arXiv:2501.07998 [gr-qc]}}.

\bibitem{Creci:2023cfx}
G.~Creci, T.~Hinderer, and J.~Steinhoff, ``{Tidal properties of neutron stars
  in scalar-tensor theories of gravity},''
  \href{http://dx.doi.org/10.1103/PhysRevD.108.124073}{{\em Phys. Rev. D}
  {\bfseries 108} no.~12, (2023) 124073},
  \href{http://arxiv.org/abs/2308.11323}{{\ttfamily arXiv:2308.11323 [gr-qc]}}.
  [Erratum: Phys.Rev.D 111, 089901 (2025)].

\bibitem{Creci:2024wfu}
G.~Creci, I.~van Gemeren, T.~Hinderer, and J.~Steinhoff, ``{Tidal effects in
  gravitational waves from neutron stars in scalar-tensor theories of
  gravity},'' \href{http://dx.doi.org/10.21468/SciPostPhysCore.8.2.042}{{\em
  SciPost Phys. Core} {\bfseries 8} (2025) 042},
  \href{http://arxiv.org/abs/2412.06620}{{\ttfamily arXiv:2412.06620 [gr-qc]}}.

\bibitem{Gannouji:2019gnb}
R.~Gannouji, Y.~Rodr\'\i{}guez~Baez, and N.~Dadhich, ``{Pure Lovelock black
  holes in dimensions $d=3N+1$ are stable},''
  \href{http://dx.doi.org/10.1103/PhysRevD.100.084011}{{\em Phys. Rev. D}
  {\bfseries 100} no.~8, (2019) 084011},
  \href{http://arxiv.org/abs/1907.09503}{{\ttfamily arXiv:1907.09503 [gr-qc]}}.

\bibitem{Dadhich:2015nua}
N.~Dadhich and J.~M. Pons, ``{Static pure Lovelock black hole solutions with
  horizon topology S$^{(n)}\times$ S$^{(n)}$},''
  \href{http://dx.doi.org/10.1007/JHEP05(2015)067}{{\em JHEP} {\bfseries 05}
  (2015) 067}, \href{http://arxiv.org/abs/1503.00974}{{\ttfamily
  arXiv:1503.00974 [gr-qc]}}.

\bibitem{Dadhich:2012cv}
N.~Dadhich, S.~G. Ghosh, and S.~Jhingan, ``{The Lovelock gravity in the
  critical spacetime dimension},''
  \href{http://dx.doi.org/10.1016/j.physletb.2012.03.084}{{\em Phys. Lett. B}
  {\bfseries 711} (2012) 196--198},
  \href{http://arxiv.org/abs/1202.4575}{{\ttfamily arXiv:1202.4575 [gr-qc]}}.

\bibitem{Gannouji:2013eka}
R.~Gannouji and N.~Dadhich, ``{Stability and existence analysis of static black
  holes in pure Lovelock theories},''
  \href{http://dx.doi.org/10.1088/0264-9381/31/16/165016}{{\em Class. Quant.
  Grav.} {\bfseries 31} (2014) 165016},
  \href{http://arxiv.org/abs/1311.4543}{{\ttfamily arXiv:1311.4543 [gr-qc]}}.

\bibitem{Dadhich:2015ivt}
N.~Dadhich, R.~Durka, N.~Merino, and O.~Miskovic, ``{Dynamical structure of
  Pure Lovelock gravity},''
  \href{http://dx.doi.org/10.1103/PhysRevD.93.064009}{{\em Phys. Rev. D}
  {\bfseries 93} no.~6, (2016) 064009},
  \href{http://arxiv.org/abs/1511.02541}{{\ttfamily arXiv:1511.02541
  [hep-th]}}.

\bibitem{Dadhich:2015lra}
N.~Dadhich, ``{A distinguishing gravitational property for gravitational
  equation in higher dimensions},''
  \href{http://dx.doi.org/10.1140/epjc/s10052-016-3933-z}{{\em Eur. Phys. J. C}
  {\bfseries 76} no.~3, (2016) 104},
  \href{http://arxiv.org/abs/1506.08764}{{\ttfamily arXiv:1506.08764 [gr-qc]}}.

\bibitem{Lovelock:1971yv}
D.~Lovelock, ``{The Einstein tensor and its generalizations},''
  \href{http://dx.doi.org/10.1063/1.1665613}{{\em J. Math. Phys.} {\bfseries
  12} (1971) 498--501}.

\bibitem{boulware1985string-generated-4e7}
D.~G. Boulware and S.~Deser, ``String-generated gravity models,''
  \href{http://dx.doi.org/10.1103/physrevlett.55.2656}{{\em Physical Review
  Letters} {\bfseries 55} no.~24, (1985) 2656--2660}.

\bibitem{garraffo2008lovelock-70a}
C.~Garraffo and G.~Giribet, ``The lovelock black holes,''
  \href{http://dx.doi.org/10.48550/arxiv.0805.3575}{{\em {arXiv}} (2008) },
  \href{http://arxiv.org/abs/0805.3575}{{\ttfamily 0805.3575}}.

\bibitem{Silvestrini:2025lbe}
M.~Silvestrini, E.~Maggio, S.~Chakraborty, and P.~Pani, ``{One Membrane to Love
  them all: Tidal deformations of compact objects from the membrane
  paradigm},'' \href{http://arxiv.org/abs/2506.16516}{{\ttfamily
  arXiv:2506.16516 [gr-qc]}}.

\bibitem{Guo:2018exx}
X.-Y. Guo, H.-F. Li, and L.-C. Zhang, ``{Entropy of higher-dimensional charged
  Gauss\textendash{}Bonnet black hole in de Sitter space},''
  \href{http://dx.doi.org/10.1088/1572-9494/ab8a25}{{\em Commun. Theor. Phys.}
  {\bfseries 72} no.~8, (2020) 085403},
  \href{http://arxiv.org/abs/1803.09456}{{\ttfamily arXiv:1803.09456 [gr-qc]}}.

\bibitem{PhysRevD.65.084014}
R.-G. Cai, ``Gauss-bonnet black holes in ads spaces,''
  \href{http://dx.doi.org/10.1103/PhysRevD.65.084014}{{\em Phys. Rev. D}
  {\bfseries 65} (Mar, 2002) 084014}.
  \url{https://link.aps.org/doi/10.1103/PhysRevD.65.084014}.

\bibitem{Cvetic:2001bk}
M.~Cvetic, S.~Nojiri, and S.~D. Odintsov, ``{Black hole thermodynamics and
  negative entropy in de Sitter and anti-de Sitter Einstein-Gauss-Bonnet
  gravity},'' \href{http://dx.doi.org/10.1016/S0550-3213(02)00075-5}{{\em Nucl.
  Phys. B} {\bfseries 628} (2002) 295--330},
  \href{http://arxiv.org/abs/hep-th/0112045}{{\ttfamily arXiv:hep-th/0112045}}.

\bibitem{Hendi:2015pda}
S.~H. Hendi, S.~Panahiyan, and B.~Eslam~Panah, ``{Charged Black Hole Solutions
  in Gauss-Bonnet-Massive Gravity},''
  \href{http://dx.doi.org/10.1007/JHEP01(2016)129}{{\em JHEP} {\bfseries 01}
  (2016) 129}, \href{http://arxiv.org/abs/1507.06563}{{\ttfamily
  arXiv:1507.06563 [hep-th]}}.

\bibitem{NIST:DLMF}
``{\it NIST Digital Library of Mathematical Functions}.''
  \url{https://dlmf.nist.gov/}, release 1.2.4 of 2025-03-15.
\newblock \url{https://dlmf.nist.gov/}. F.~W.~J. Olver, A.~B. {Olde Daalhuis},
  D.~W. Lozier, B.~I. Schneider, R.~F. Boisvert, C.~W. Clark, B.~R. Miller,
  B.~V. Saunders, H.~S. Cohl, and M.~A. McClain, eds.

\end{thebibliography}\endgroup

\bibliographystyle{./utphys1}

\end{document}